\documentclass[twocolumn]{aastex61}
\usepackage{amsmath,amstext}
\usepackage[T1]{fontenc}
\usepackage{apjfonts} 
\usepackage[figure,figure*]{hypcap}
\usepackage{commath}

   \newcommand{\rah}{$^{\mbox{\scriptsize h}}$}
   \newcommand{\ram}{$^{\mbox{\scriptsize m}}$}
   \newcommand{\ras}{$^{\mbox{\scriptsize s}}$}
   \newcommand{\decd}{$^{\circ}$}
   \newcommand{\decm}{$'$}
   \newcommand{\decs}{$''$}
   
   \newcommand{\beam}{$\theta_{\mbox{\scriptsize maj}}\times\theta_{\mbox{\scriptsize min}}$}
   \newcommand{\mjyperbeam}{mJy\,beam$^{-1}$}
   \newcommand{\ujyperbeam}{$\mu$Jy\,beam$^{-1}$}
   
   \newcommand{\vlsr}{$v_{\mbox{\scriptsize lsr}}$}
   
   \newcommand{\nraojansky}{\affiliation{\it{Jansky fellow of the National Radio Astronomy Observatory, 1003 Lopezville Rd, Socorro, NM 87801 USA }}}

\shorttitle{Toomre instability assisted fragmentation in OB cluster-forming molecular clumps}
\shortauthors{Liu et al.}

\begin{document}

\title{Investigating fragmentation of gas structures in OB cluster-forming molecular clump G33.92+0.11 with 1000 AU resolution observations of ALMA}


\correspondingauthor{Hauyu Baobab Liu}
\email{baobabyoo@gmail.com}

\author[0000-0003-2300-2626]{Hauyu Baobab Liu}
\affil{
Academia Sinica Institute of Astronomy and Astrophysics, P.O. Box 23-141, Taipei 10617, Taiwan
}
\affiliation{European Southern Observatory (ESO), Karl-Schwarzschild-Str. 2, 85748, Garching, Germany} 

\author[0000-0002-9774-1846]{Huei-Ru Vivien Chen}
\affil{Institute of Astronomy and Department of Physics, National Tsing Hua University, Hsinchu 30013, Taiwan}

\author[0000-0001-8600-4798]{Carlos G. Rom{\'a}n-Z{\'u}{\~n}iga}
\affil{
(Instituto de Astronom\'ia, Universidad Nacional Autónoma de M\'exico, Apartado Postal 106, 22800 Ensenada, Baja California, Mexico
}

\author[0000-0002-0786-7307]{Roberto Galv\'an-Madrid}
\affil{
Instituto de Radioastronom\'ia y Astrof\'isica, Universidad Nacional Aut\'onoma de M\'exico, Apdo. Postal 72-3 (Xangari), Morelia, Michoac\'an 58089, M\'exico. 
}

\author[0000-0001-6431-9633]{Adam Ginsburg}
\nraojansky

\author{Paul T.~P. Ho}
\affil{
Academia Sinica Institute of Astronomy and Astrophysics, P.O. Box 23-141, Taipei 10617, Taiwan
}
\affil{
East Asian Observatory, 666 N. A'ohoku Place, Hilo, Hawaii, HI 96720, USA
}

\author[0000-0003-1742-0119]{Young Chol Minh}
\affil{
Korea Astronomy and Space Science Institute, 776 Daedeok-daero, Yuseong, Daejeon 34055, Korea
}

\author[0000-0003-4493-8714]{Izaskun Jim\'enez-Serra}
\affil{
School of Physics and Astronomy, Queen Mary University of London, Mile End Road E1 4NS, London, UK
}

\author[0000-0003-1859-3070]{Leonardo Testi}
\affil{
European Southern Observatory (ESO), Karl-Schwarzschild-Str. 2, 85748, Garching, Germany
}

\author[0000-0003-2384-6589]{Qizhou Zhang}
\affil{
Harvard-Smithsonian Center for Astrophysics, 60 Garden Street, Cambridge MA 02138, USA
}

\begin{abstract}
We report new, $\sim$1000 AU spatial resolution observations of 225 GHz dust continuum emission towards the OB cluster-forming molecular clump G33.92+0.11.
On parsec scales, this molecular clump presents a morphology with several arm-like dense gas structures surrounding the two central massive ($\gtrsim$100 $M_{\odot}$) cores.
From the new, higher resolution observations, we identified 28 localized, spatially compact dust continuum emission sources, which may be candidates of young stellar objects.
Only one of them is not embedded within known arm-like (or elongated) dense gas structures.
The spatial separations of these compact sources can be very well explained by Jeans lengths.
We found that G33.92+0.11 may be consistently described by a marginally centrifugally supported, Toomre unstable accretion flow which is approximately in a face-on projection.
The arm-like overdensities are natural consequence of the Toomre instability, which can fragment to form young stellar objects in shorter time scales than the timescale of the global clump contraction.
On our resolved spatial scales, there is not yet evidence that the fragmentation is halted by turbulence, magnetic field, or stellar feedback.
\end{abstract}


\section{Introduction}\label{sec:introduction}

Molecular clouds may undergo global collapse \citep[e.g.,][and references therein]{Bate2003,Hartmann2012,Enrique2017}, which may lead to a centrally concentrated density distribution of gas and young stellar objects (YSOs) \citep[e.g.,][and references therein]{Liu2012a,Galvan2013,Lin2017}.
As a consequence of the accumulated angular momentum, the central flattened rotating gas clump in the collapsing molecular cloud may present spin-up motions even on parsec scales \citep[e.g.,][]{Ho1986,Keto1987,Welch1987,Ho1996,Zhang1997,Galvan2009,Liu2010}.
The central cluster-forming clump may be marginally supported by rotational motion, until the collected gas mass is sufficient to trigger the self-gravitational instability, which will then result in spiral arm-like dense gas structures and a distribution of localized gas fragments \citep[e.g.,][and references therein]{Keto1991,Lee2016a,Lee2016b,Sakurai2016,Mapelli2017}.
Such structures may have been resolved in some previous observations \citep[e.g.,][]{Liu2012b,Takahashi2012,Beuther2013,Wright2014,Liu2015,Chen2016,Li2017,Beuther2018,Izquierdo2018,Maud2017}.
How these gas structures fragment to subsequently form 10$^{3}$-10$^{4}$ AU scales gas cores and YSOs, remain uncertain.

To well resolve the gas structures forming out of the self-gravitational fragmentation in the centralized massive molecular gas clumps in OB cluster-forming molecular clouds, we selected to observe the target source G33.92+0.11 ($d\sim$7.1 kpc).
Its very small derived virial mass compared to the enclosed molecular gas mass \citep[][]{Watt1999} indicates that it is likely geometrically thin and is in a face-on projection.
It encloses a few thousands $M_{\odot}$ of gas mass in the central parsec scale area \citep[][]{Liu2012b}.
However, the previous interferometric observations of molecular lines resolved very small relative motions with respect to its systemic velocity 107.6 km\,s$^{-1}$ \citep[e.g., within $\pm$2 km\,s$^{-1}$ for most of the regions. For more details see][]{Liu2015,Minh2016}.
This was interpreted as motions predominantly in the plane of the sky, with the dominant motion being rotation, where the axis of rotation is parallel to the line of sight.  
If this is indeed the case, then the studies of its matter distribution will be minimally affected by line-of-sight confusion.

We have resolved G33.92+0.11 using the Atacama Large Millimeter Array (ALMA) and Atacama Compact Array (ACA), with a $\sim$1000 AU spatial resolution.
The observations and data reduction are introduced in Section \ref{sec:observation}.
The direct observational results are presented in Section \ref{sec:results}.
In Section \ref{sub:Toomre} we address the gravitational instability of the resolved system based on the analysis of the Toomre Q parameter.
In Section \ref{sub:dendrogram} we present our identification of the localized candidates of YSOs using the {\tt dendrogram} algorithm \citep{Rosolowsky2008}; and in Section \ref{sub:cluster} and \ref{sub:separation} we discuss the clustering of the identified YSO candidates, and the probable physical mechanism to explain their spatial separations.
Our conclusion is given in Section \ref{sec:conclusion}.

\begin{figure*}
  \hspace{0.3cm}
  \begin{tabular}{ p{8cm} p{8cm} }
    \multicolumn{2}{l}{
      \includegraphics[width=14cm]{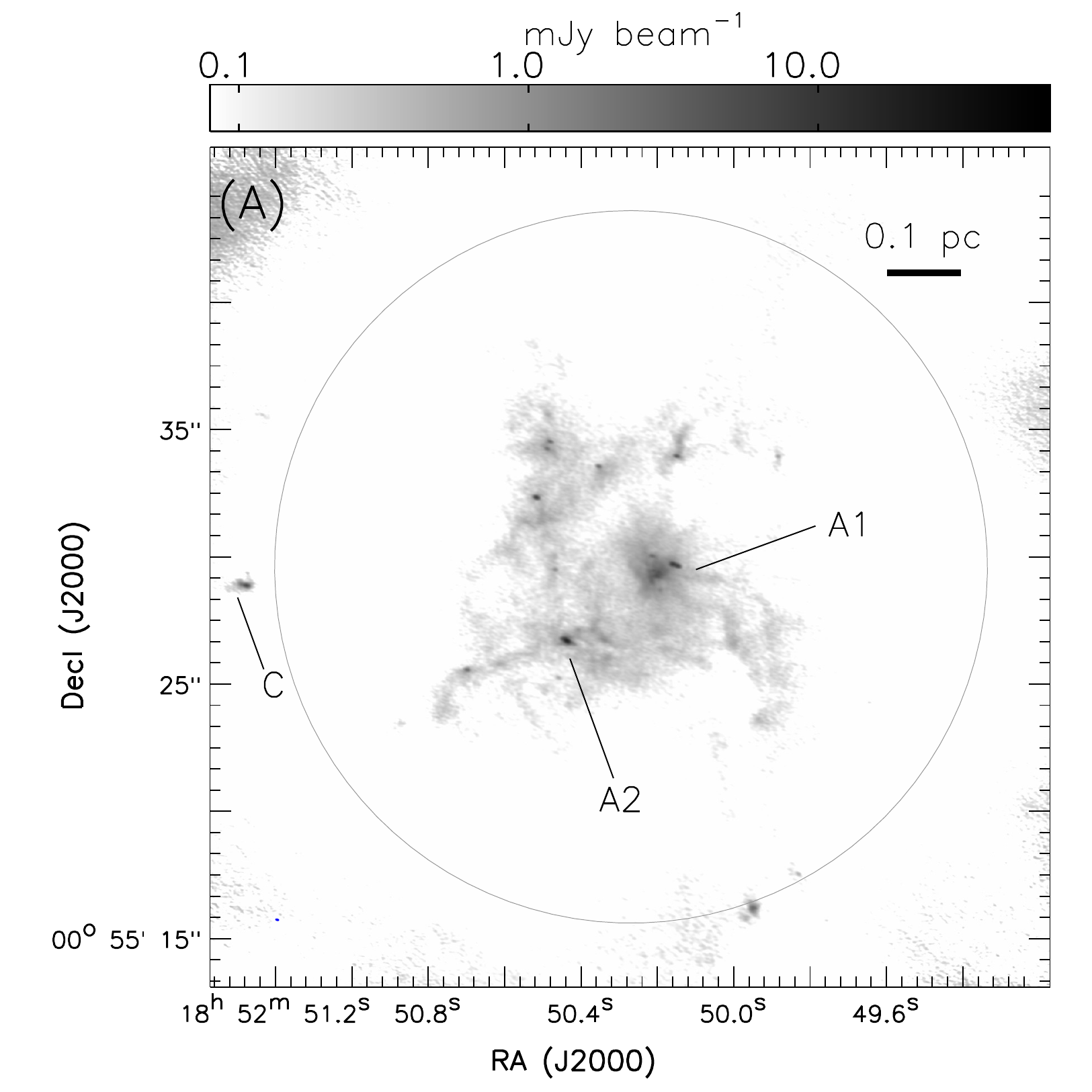} 
    }\\
    \includegraphics[width=7.5cm]{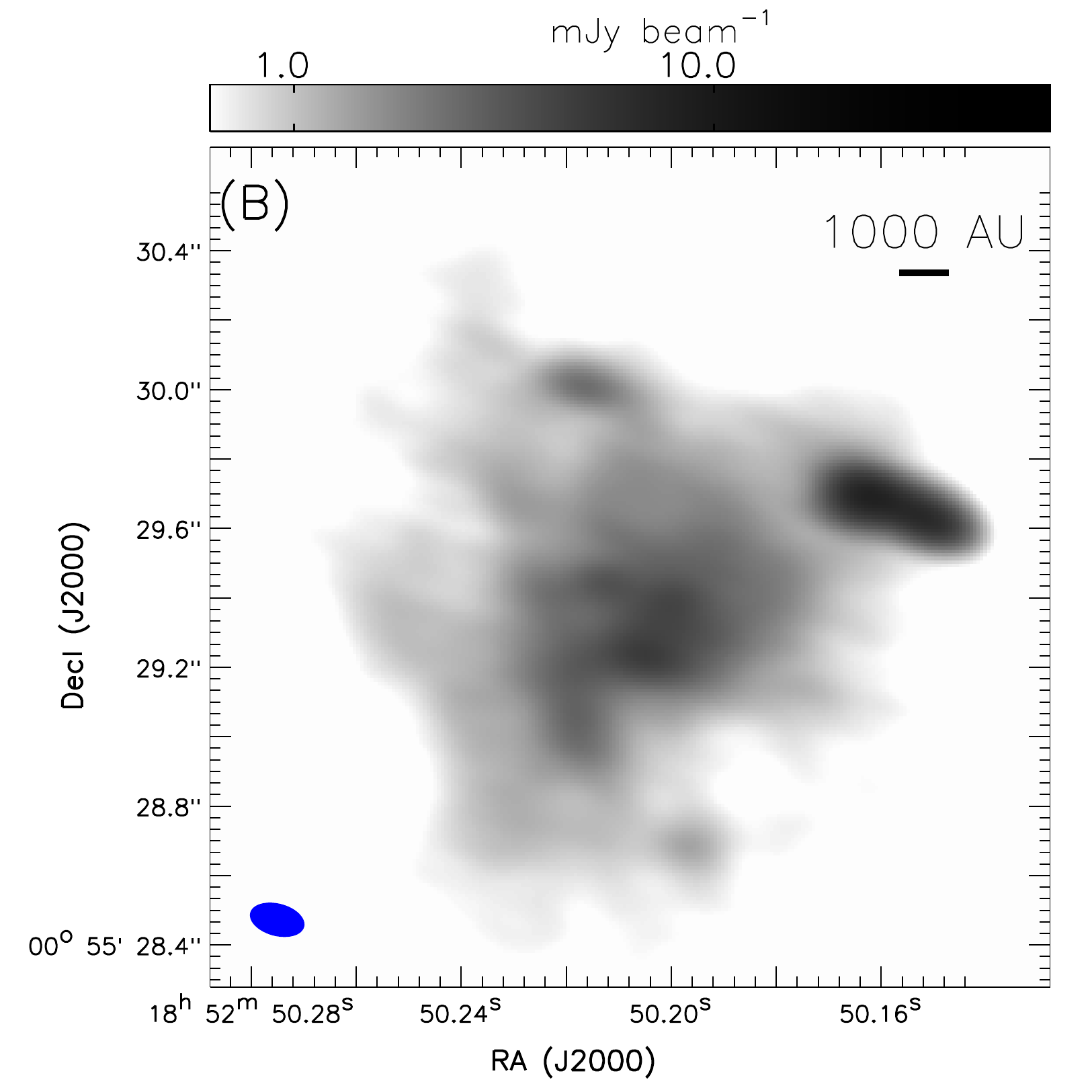} &
    \includegraphics[width=7.5cm]{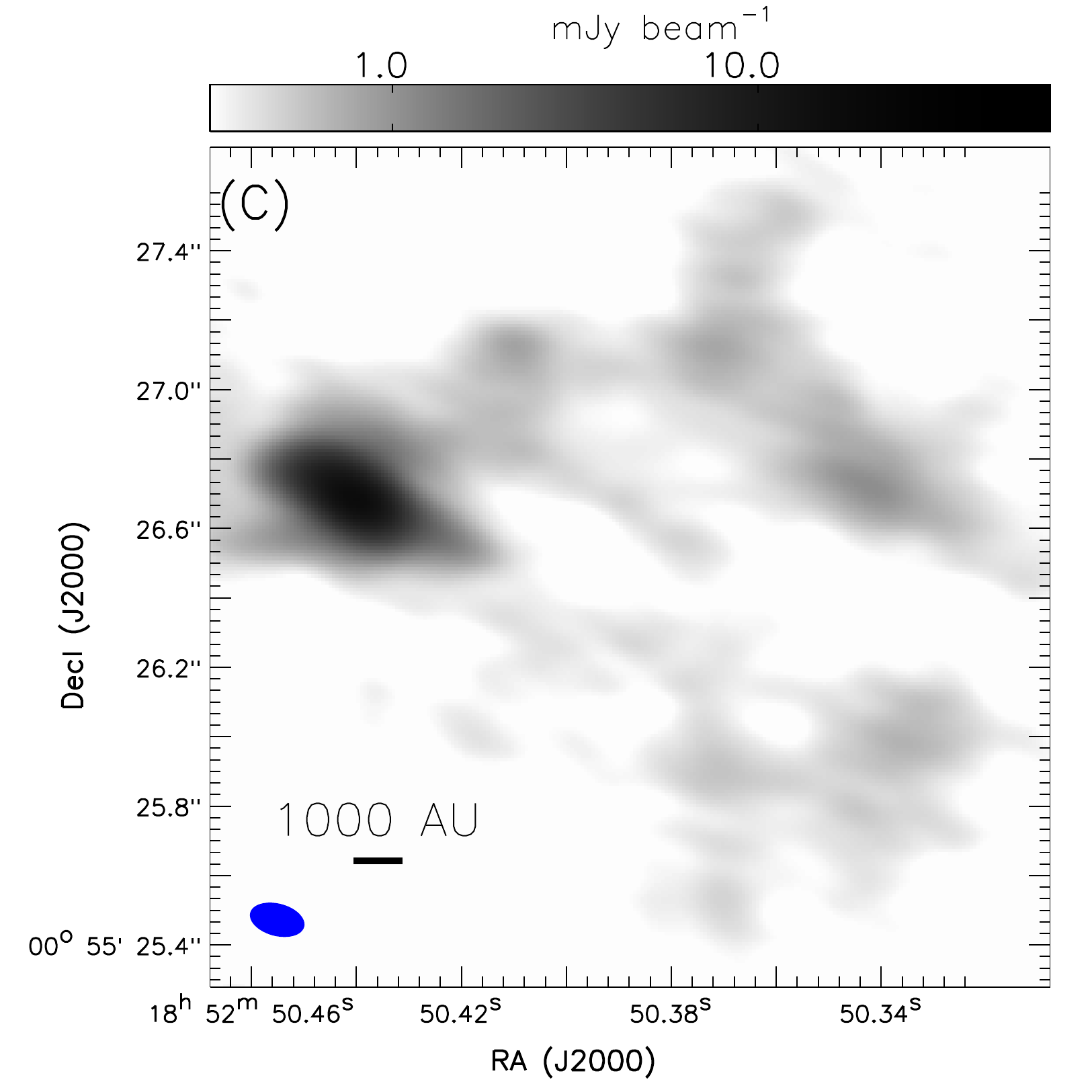} \\
  \end{tabular}
  \vspace{0cm}
  \caption{
     The 1.3 mm continuum image generated by combining all existing ALMA and ACA data. The full width of half maximum (FWHM) of the primary beam is 28$''$, as indicated with the gray circle. Blue ellipses show the synthesized beam (\beam = 0$''$.16$\times$0$''$.093; P.A.=74$^{\circ}$). Panel (A), (B) and (C) show respectively the full image, and the smaller regions around the A1 and A2 {\it central massive  molecular cores}.
  }
  \label{fig:almacycle4}
  \vspace{-0.55cm}
\end{figure*}

\begin{figure}
  \includegraphics[width=8.5cm]{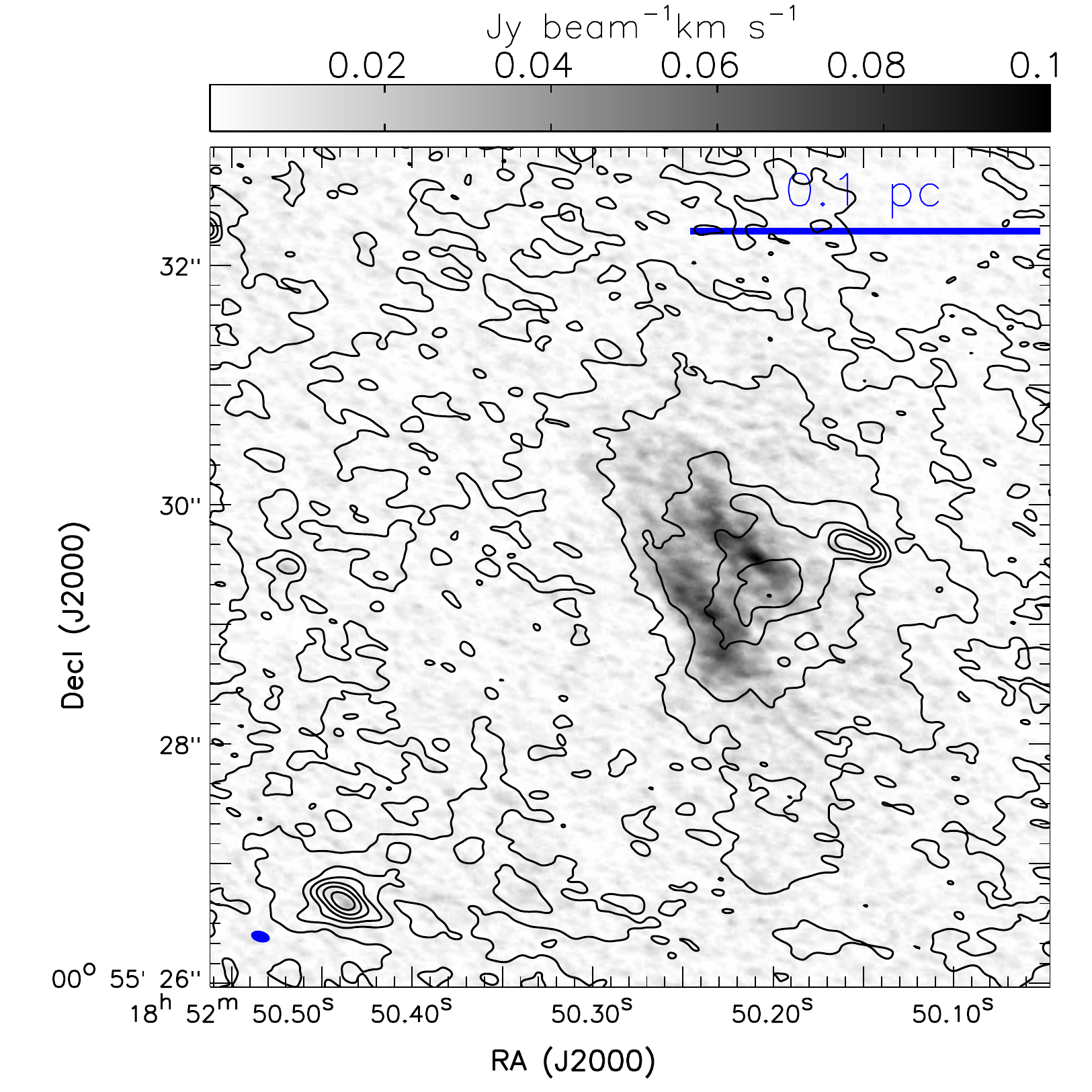}
  \caption{The velocity integrated intensity map (i.e., moment 0) of the H30$\alpha$ line (gray), overlaid with the 1.3 mm continuum image is presented in Figure \ref{fig:almacycle4}. The two images share the same synthesized beam. Contour levels are  25 $\mu$Jy\,beam$^{-1}$ $\times$ [3, 6, 12, 24, 48, 96, 192, 384, 768].
  \label{fig:h30alpha}
  }
\end{figure}

\section{Observation and data reduction}\label{sec:observation}
We have performed ALMA and ACA observations towards G33.92+0.11 (Project codes: 2012.1.00387.S, 2016.1.00362.S, PI: H. B. Liu, H.-R. Chen).
The pointing and phase referencing center is R.A. (J2000) = 18\rah52\ram50\ras.272, and Decl. (J2000) = 00\decd55\decm29\decs.604.
The spectral setup of all of our observations are identical.
There were two 234.4 MHz wide spectral windows (channel spacing 61 kHz, $\sim$0.085 km\,s$^{-1}$) centered at 231.220690 GHz and 220.679320 GHz, and two 1875 MHz wide spectral windows (channel spacing 488 kHz, $\sim$0.65 km\,s$^{-1}$) centered at 231.900928 GHz and 217.104980 GHz. 
These spectral windows tracked the systemic velocity \vlsr$\sim$107.6 km\,s$^{-1}$.

\begin{figure*}
  \hspace{-1.5cm}
  \vspace{0cm}
  \begin{tabular}{ p{9.2cm} p{9.2cm} }
    \vspace{-5cm}\includegraphics[width=10.5cm]{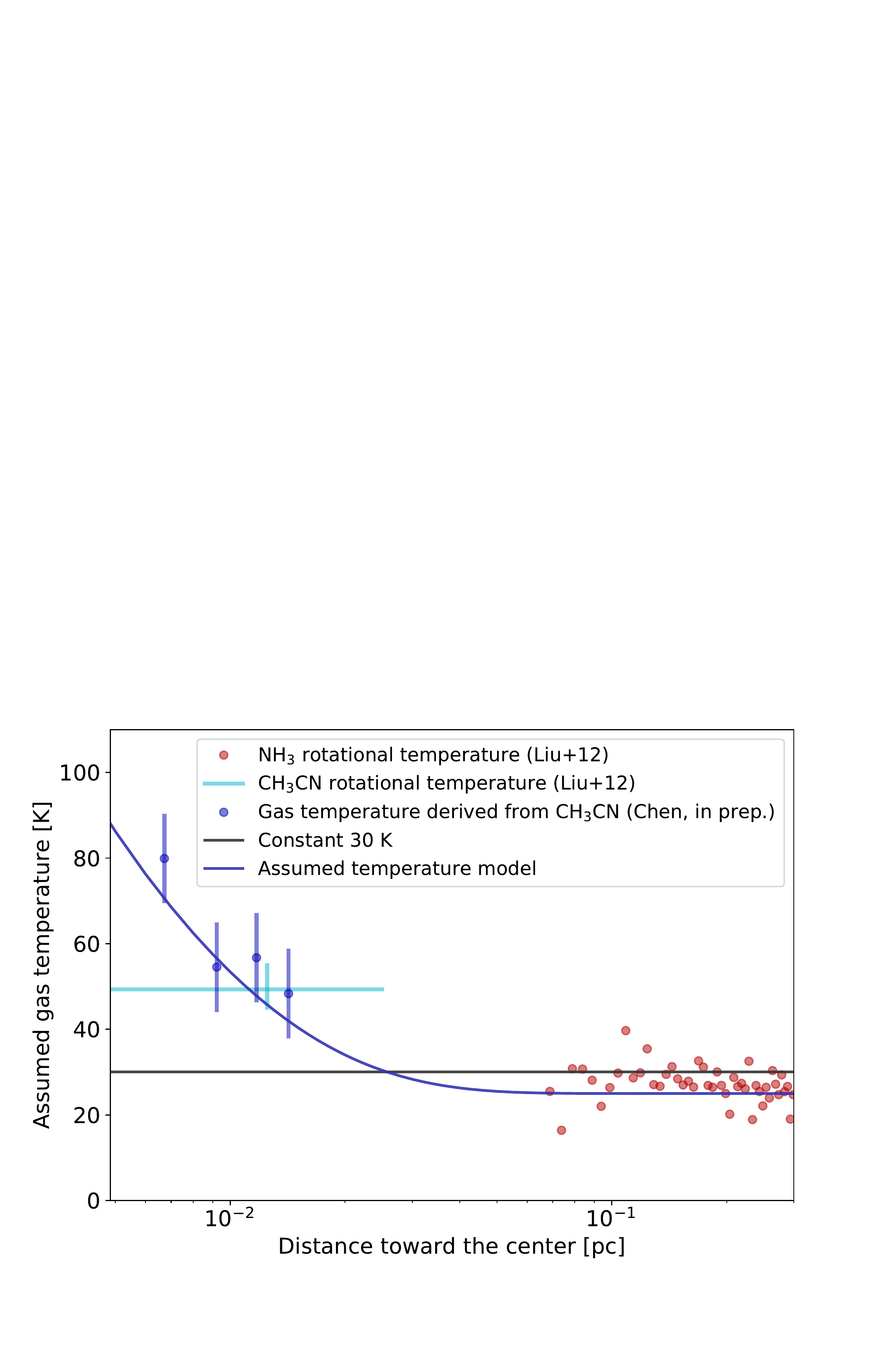} &
    \vspace{-5cm}\includegraphics[width=10.5cm]{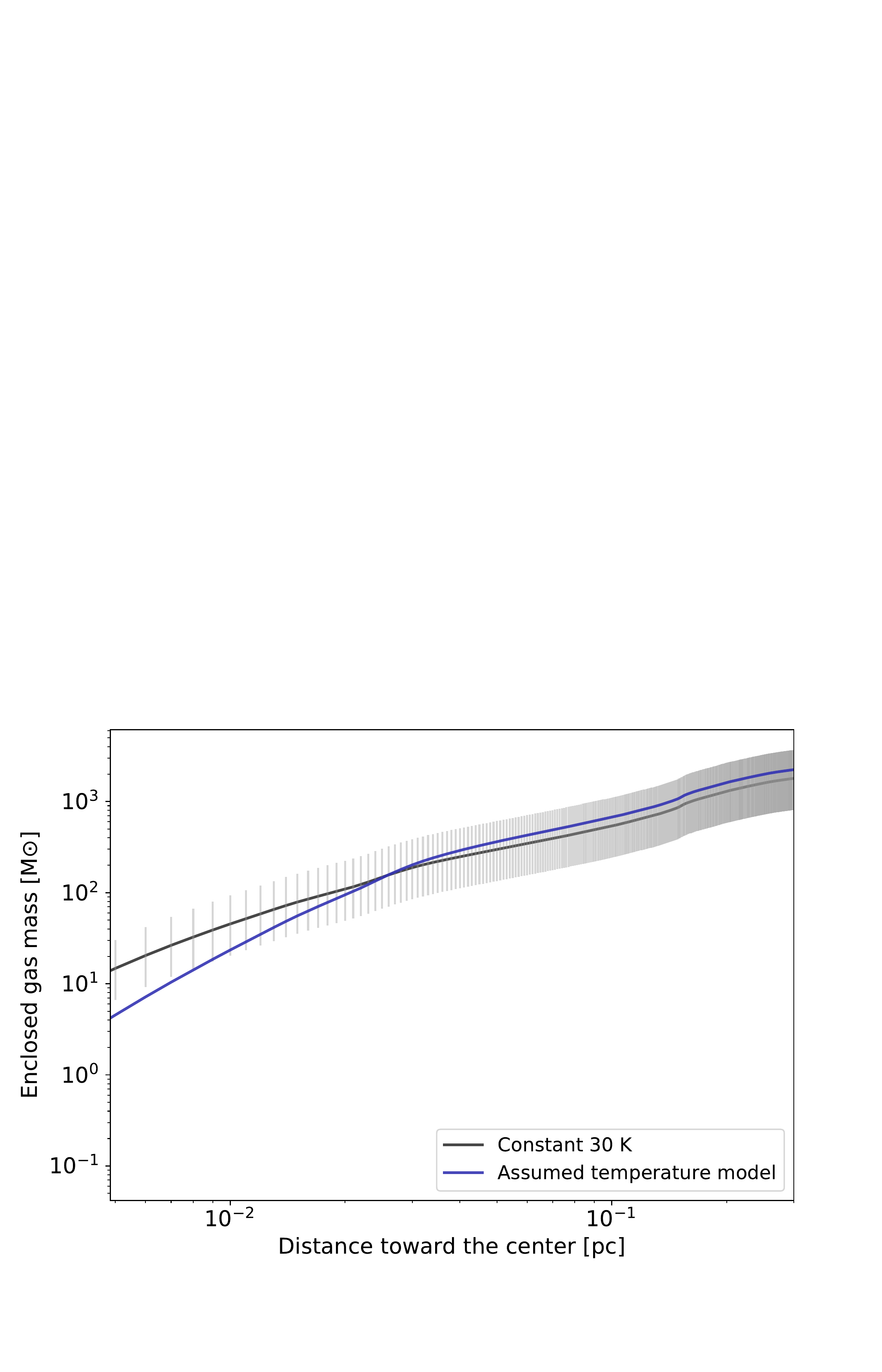} \\
  \end{tabular}
    
  \vspace{-9.9cm}\hspace{-1.5cm}
  \begin{tabular}{ p{9.2cm} p{9.2cm} }
    \vspace{0cm}\includegraphics[width=10.5cm]{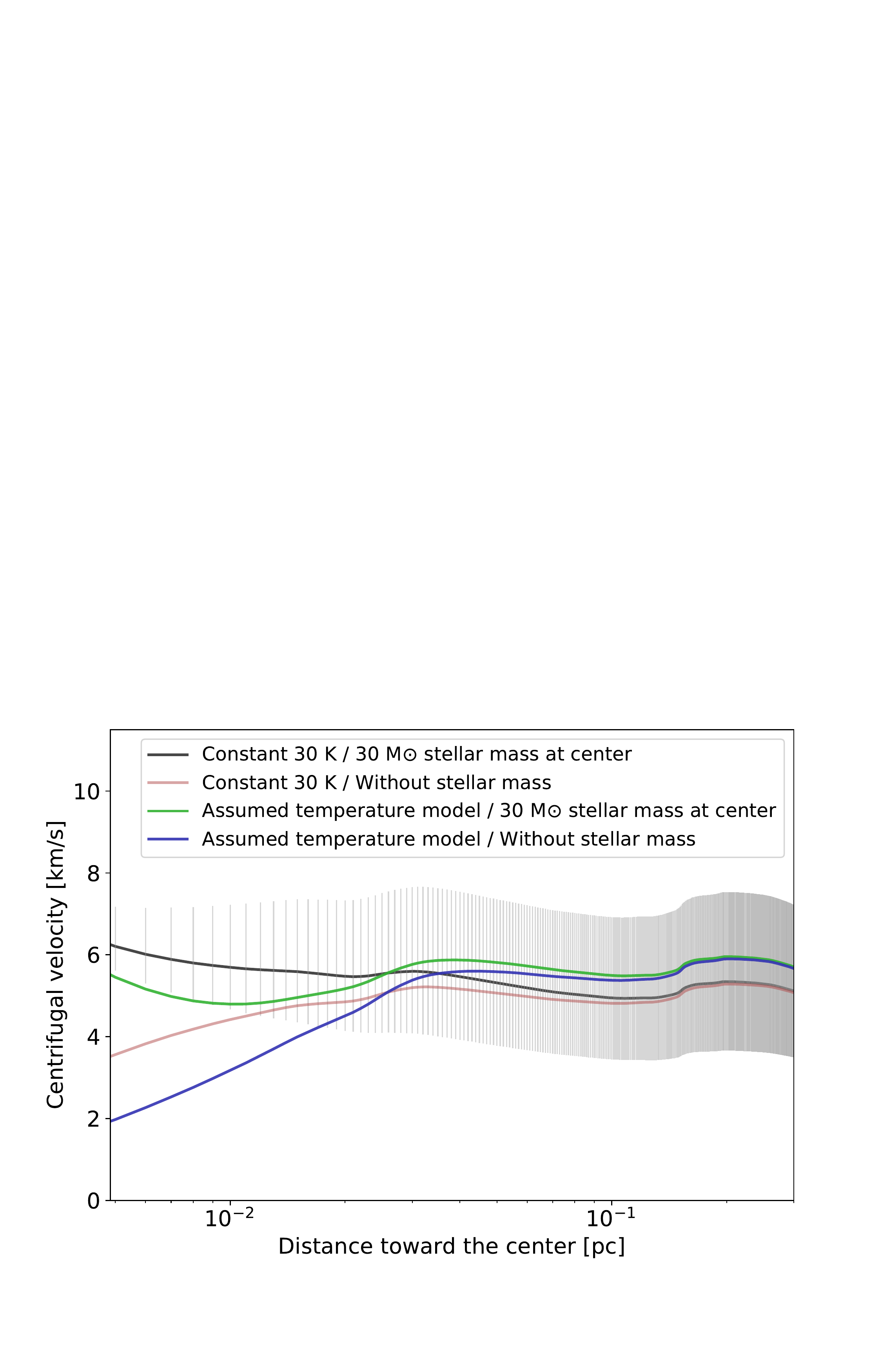} &
    \vspace{0cm}\includegraphics[width=10.5cm]{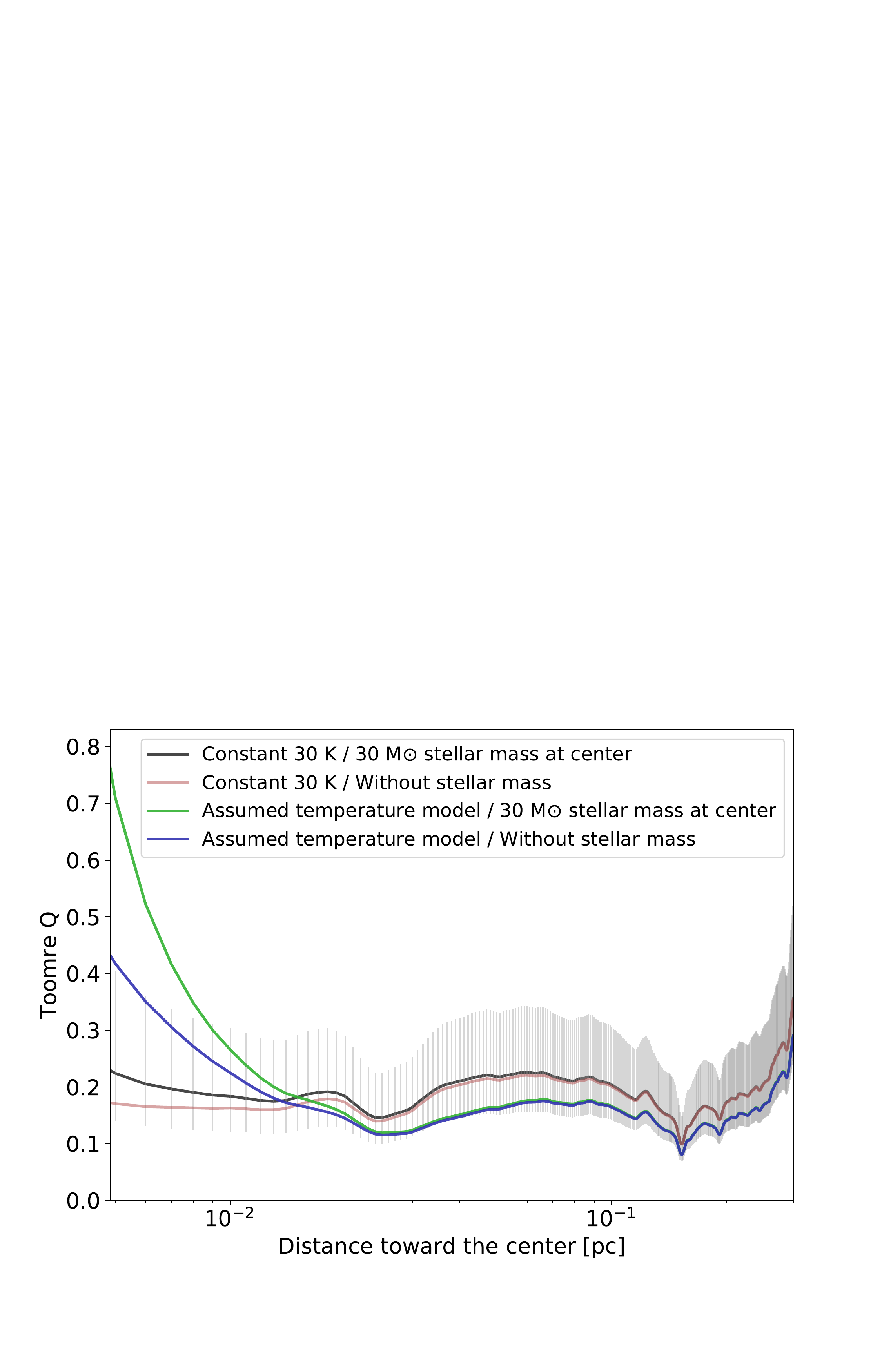} \\
  \end{tabular}
  \vspace{-1.3cm}
  \caption{
    The assumed temperature profile, compared with the measurements reported by \citet{Liu2012b} ({\it top left}), the enclosed gas mass ({\it top right}; assuming dust mass opacity $\kappa_{\mbox{\tiny 225 GHz}}$=0.6 cm$^{2}$\,g$^{-1}$), the derived centrifugal velocity ({\it bottom left}), and the Toomre Q parameter ({\it bottom right}). Error bars characterize the $\kappa_{\mbox{\tiny 225 GHz}}$ range of 0.3-1.3 cm$^{2}$\,g$^{-1}$, and are only presented for the case where the assumed stellar mass is included (and only for the case assuming constant 30 K temperature in the bottom panels). The error bars for the other cases are similar to the presented ones. The presented smallest radius (i.e., distance toward the center) corresponds to our spatial resolution ($\sim$1000 AU). In the top left panel we also present the gas temperature inward of the 3000 AU radius probed by the high angular resolution CH$_{3}$CN J=12-11 data, which are simultaneously covered by our observations introduced in Section \ref{sec:observation}. However, we caution that the inner $\sim$0.1 pc region is a complicated, cluster-forming environment, and that the assumption of a simple radial temperature profile is not generally validate (Chen et al. in prep.) 
          }
  \label{fig:Toomre}
  \vspace{-0cm}
\end{figure*}

The ALMA observations on 2014 May 04, and the ACA observations on 2014 May 03 and 04, were reported by \citet{Liu2015}.
There was an epoch of ALMA observations on 2014 April 29, which had failed the official quality assurance (QA2) test, and that was not analyzed previously.
We manually calibrated these observations and found that the data can be used, and thus included them in our present analysis.
The achieved synthesized beam (Briggs Robust = 0 weight) by these previous ALMA and ACA data is \beam = 0\farcs67$\times$0\farcs47.
We carried out the new, long baseline array configuration observations of ALMA on 2017 August 19 (UTC 01:54-04:31), aiming to achieve an angular resolution of 0\farcs1.
The {\it uv} distance range covered by the 2017 observations is $\sim$18-3400 meters (i.e., $\sim$13.5-2540 $k\lambda$ at the averaged frequency $\sim$224.5 GHz).

Combining all existing data yields an overall {\it uv} distance range of 7-3400 meters.
The data were calibrated and phase self-calibrated using the CASA software package \citep{McMullin2007} version 5.0.0.
We fit the continuum baselines using the CASA task {\tt uvcontsub}, and then jointly imaged all continuum data using the CASA task {\tt clean}.
Images are created with 6000 pixels in the R.A. and Decl. dimensions, with 0\farcs01 pixel size.
To achieve a high intensity dynamic range, the effect of the spectral index distribution is not negligible.
However, there are spatially extended and low signal-to-noise structures distributed across our field of view (FOV).
We found that in this case, the convergence of the multi-frequency synthesis (MFS) imaging with more than one Taylor terms is not robust.
Therefore, we performed MFS imaging for the upper and lower sidebands separately using only one Taylor term, and then linearly combined the upper and lower sideband images after smoothing to an identical angular resolution.
As restoring beam we used Briggs Robust = 0 weighted continuum image achieved a synthesized beam of \beam = 0\farcs16$\times$0\farcs093 ($\sim$1100 AU $\times$660 AU; P.A.=75$^{\circ}$).
The root mean square (RMS) noise estimated based on the difference of the upper and lower sidebands images is 25 \ujyperbeam ($\sim$41 mK), and the peak intensity in the primary beam corrected image is 17 \mjyperbeam ($\sim$28 K).
Details about the simultaneously covered spectral lines will be introduced in the forthcoming paper(s) (Minh et al. submitted).

\begin{figure*}
  \vspace{1cm}
  \hspace{-1.5cm}
  \begin{tabular}{ p{8.9cm} p{8.9cm} }
    \includegraphics[width=9.7cm]{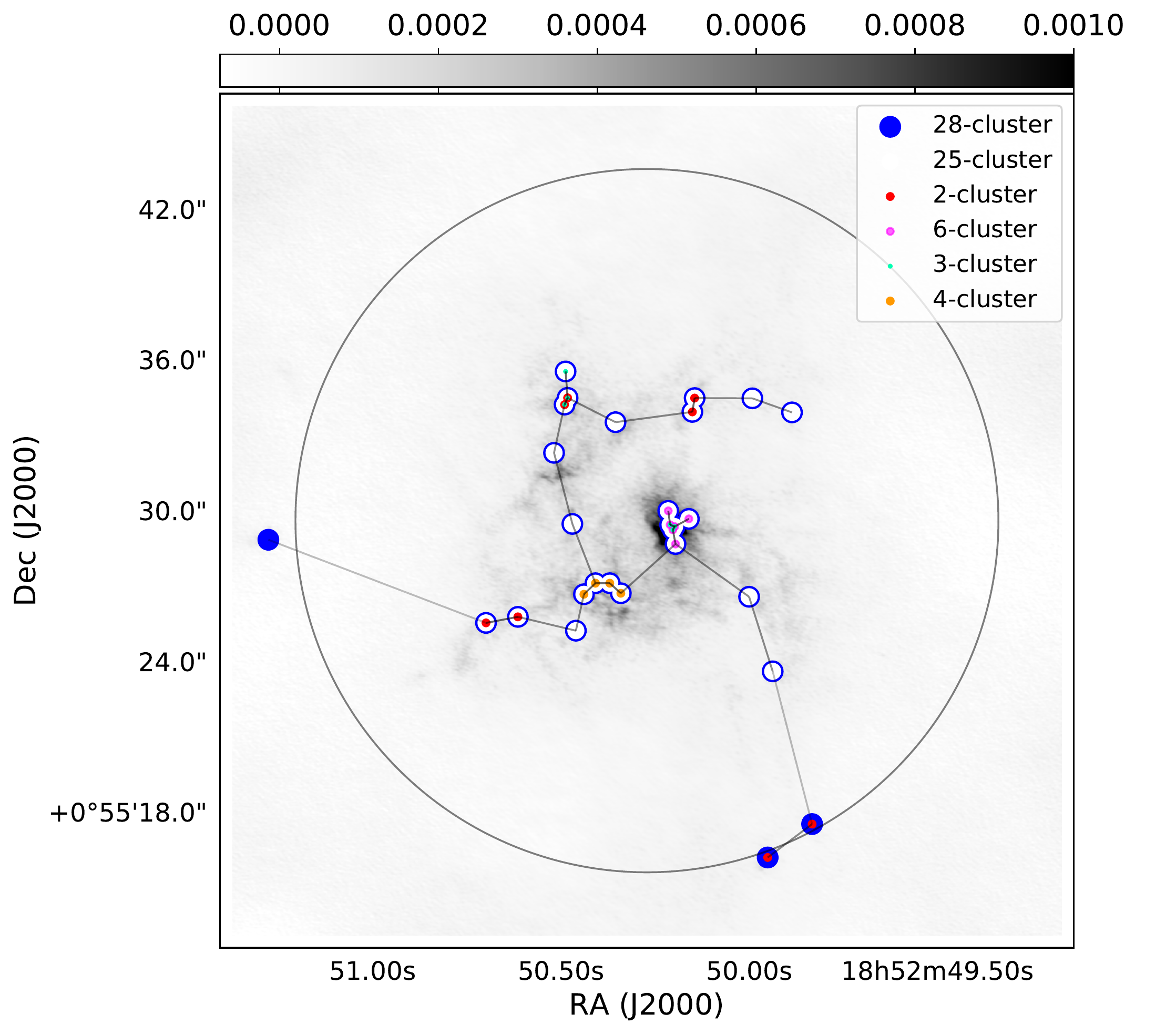} &
    \includegraphics[width=9.7cm]{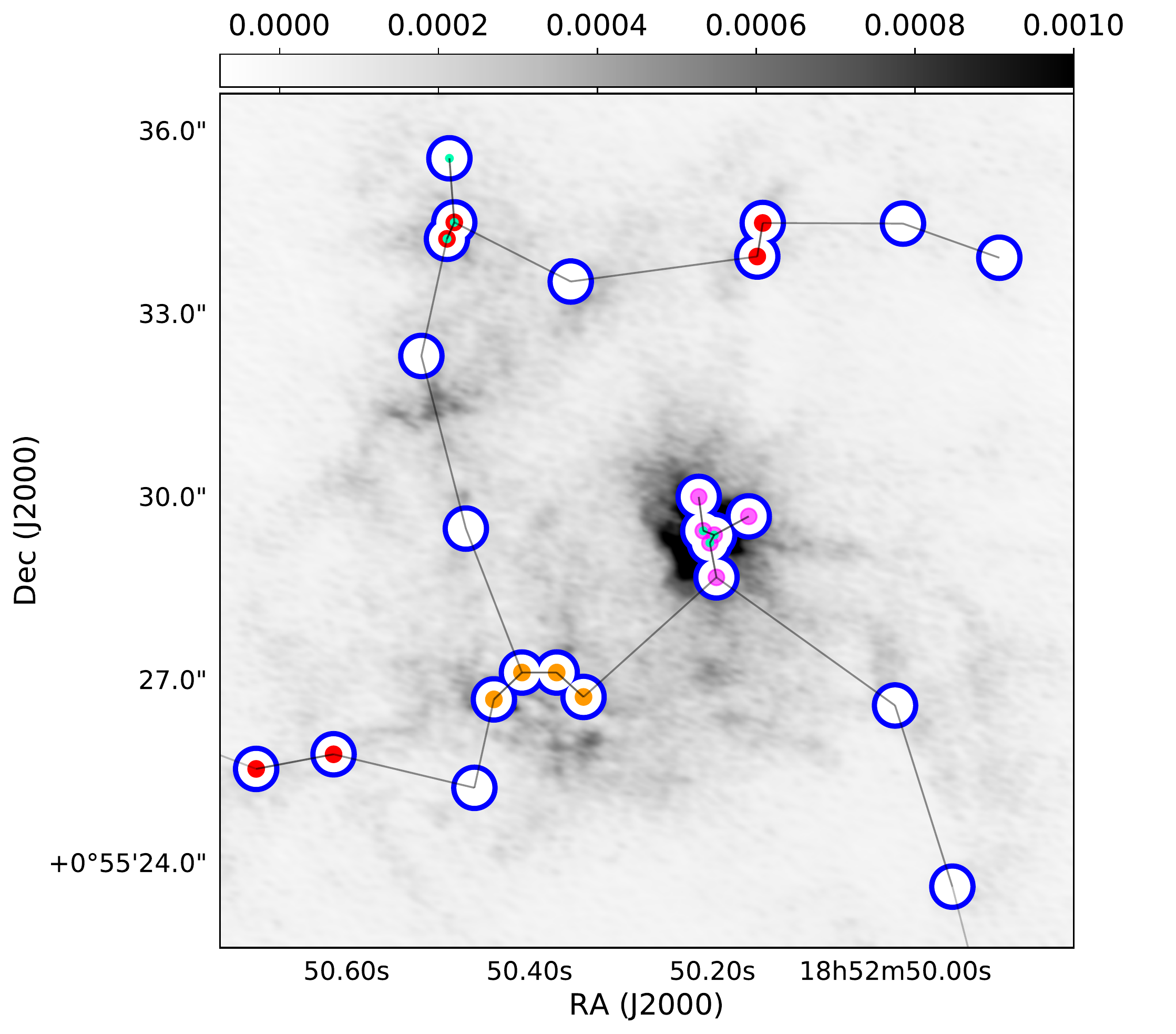} \\
  \end{tabular}
  \caption{
    The identified candidates of young stellar objects (circles), overplotted on the ALMA continuum image (grayscale). Left and right panels show the full field-of-view and the center region, respectively. The candidates of young stellar objects are assigned as members of {\it n}-cluster (Section \ref{sub:cluster}), which are color coded for different values of {\it n}. Line segments are the links in between the members of the 28-cluster. The color coding are presented in the legend of the left panel. Color bars are in units of Jy\,beam$^{-1}$ (\beam=0\farcs16$\times$0\farcs093; P.A.=74$^{\circ}$). The original output from {\tt dendrogram} is provided in Appendix \ref{appendix:dendrogram}.
          }
  \label{fig:cluster}
\end{figure*}

\begin{figure}
    \hspace{-0.6cm}
    \vspace{0cm}
    \includegraphics[width=9.5cm]{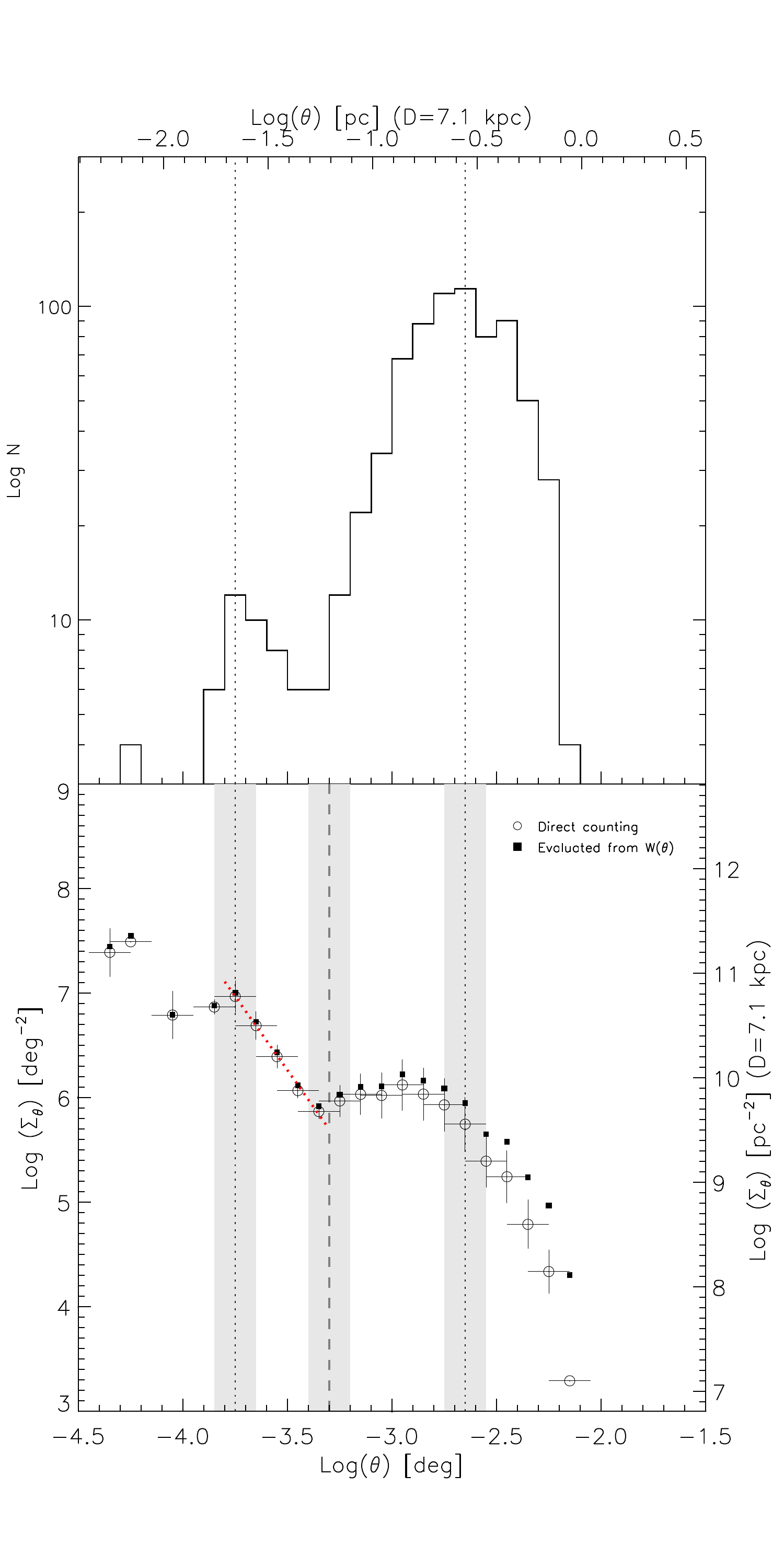} 
  \caption{
    The top panel shows the distribution of the number of
  companions as a function of the separation, $\theta$ in logarithmic space. 
  The bottom panel shows the MSDC estimated both from direct counts 
  (open symbols) and from the TPCF (solid symbols). The red, dotted line
  indicates a linear fit to the points in the $-3.75<\log{\theta}<-3.3$
  interval. The dotted and dashed lines delimit the regions discussed
  in the text near 4500, 12800 and 57100 AU. For more details see Section \ref{sub:cluster}.
          }
  \label{fig:MSDC}
\end{figure}

\section{Results}\label{sec:results}
Figure \ref{fig:almacycle4} shows the Briggs Robust = 0 weighted continuum image.
We note that outside of the primary beam full width of half maximum (FHWM) of the 12m dish of ALMA, it is not trivial to weight in between the ALMA and ACA data, and therefore the image cannot be correctly synthesized with the software we are presently using.
Within the primary beam full width of half maximum (FWHM) of the 12m dish (28$''$), our presently achieved RMS noise level and synthesized beam area are improved to be 8 times and 21 times smaller than the continuum image published previously by \citet{Liu2015}.
Assuming the dust temperature $T_{\mbox{\scriptsize dust}}=$30 K, dust mass opacity $\kappa_{\mbox{\scriptsize 225 GHz}}=$0.6$^{+0.7}_{-0.3}$ \citep{Draine2006}, gas to dust mass ratio 100, the RMS noise level (i.e., 1-$\sigma$) of this image corresponds to a gas mass of 0.026$^{+0.026}_{-0.014}$ $M_{\odot}$.
Without considering the confusion limit, our presently achieved 3-$\sigma$ sensitivity and the achieved spatial resolution (Section \ref{sec:observation}) is sufficient for detecting the relatively bright population of low-mass Class 0/I YSOs \citep[e.g.,][]{Jes2007,Maury2010,Chen2013,LiLiu2017,Pokhrel2018}.
For the sake of clarity, in the following discussion we refer to candidates of embedded (by circumstellar envelope) young stellar objects as Class 0/I candidates, regardless of whether the sources are (presently) massive or not.

We generated a model of free-free emission image based on the hydrogen recombination line H30$\alpha$ image  cube simultaneously covered by the ALMA observations introduced in Section \ref{sec:observation}, and the following assumption of the peak line-to-continuum intensity ratio \citep{Condon2016}:
\[
7.0\cdot10^3\left(\frac{\delta v}{km\,s^{-1}}\right)^{-1}\left(\frac{\nu}{GHz}\right)^{1.1}\left(\frac{T_{e}}{K}\right)^{-1.15}\left(1+\frac{N(He^{+})}{N(H^{+})}\right)^{-1},
\]
where $\delta v$ is the FWHM of the hydrogen radio recombination line, $\nu$ is the observing frequency, and $T_{e}$ is the electron temperature, and $N(He^{+})/N(H^{+})$ is the $He^{+}$ to $H^{+}$ ion ratio which is adopted to be 0.08. 
The velocity integrated intensity (i.e., moment 0) map of the H30$\alpha$ line is provided in Figure \ref{fig:h30alpha}.
We adopted the nominal value of $T_{e}\sim$8000 K for the electron temperature \citep[e.g.,][]{Keto2008}, and estimated $\delta v$ based on the
observed velocity dispersion from the H30$\alpha$ line image cube.
We smoothed our model of free-free emission to the same angular resolution of our ALMA continuum image, and then subtracted the model from the continuum image to yield a dust continuum image.
Similar to what was suggested by \citet{Liu2015} based on the lower angular resolution observations, we found that free-free emission is likely faint, such that subtracting the free-free emission or not does not significantly impact our analysis.

We are surprised that, visually, most of the over-intensities seen from the better sensitivity and higher angular resolution dust continuum image were also seen in the previous observations of \citet{Liu2015}.
However, some of those over-intensities are now resolved with internal sub-structures.
A similar phenomenon was also reported by the previous ALMA case study on the infrared dark cloud G28.34+0.06 \citep{Zhang2015}.

For example, the previously identified, central massive ($\gtrsim$100 $M_{\odot}$) cores A1 and A2 \citep{Liu2012b,Liu2015}, now appear like (sub-)clusters of gas overdensities.
The $\sim$0.3 pc scale arm-like gas structures around A1 and A2 embed spatially unresolved compact continuum sources, which may be regarded as Class 0/I candidates.
Many of them are known to be associated with SiO jets \citep[][see also, Appendix \ref{appendix:sio}]{Minh2016}.
The overall morphology of the dense gas structures appears hierarchical.
The spatially unresolved compact dust continuum sources, which may be Class 0/I candidates, are linked with elongated structures with few thousands AU scales, which are the denser parts of the gas structures on further larger scales.
There appears to be no isolated, $\sim$0.1 pc scale dense cores, in the high angular resolution image, except for the compact source, namely source C, which is located around the eastern edge of our 12m dish primary beam.
Besides source C, there appears to be no Class 0/I candidates forming outside of the known arm-like dense gas structures.
The overall morphology of G33.92+0.11 may be qualitatively similar to the hydrodynamic simulations of gravitationally unstable, rotating high mass star-forming clump presented by \citet{Sakurai2016}.
In the following sections we test whether or not the resolved morphology is indeed consistent with a self-gravitationally unstable accretion flow, and discuss the necessary caveats which may be tested by future observations.

\begin{figure}
    \vspace{-1.2cm}
    \includegraphics[width=9.7cm]{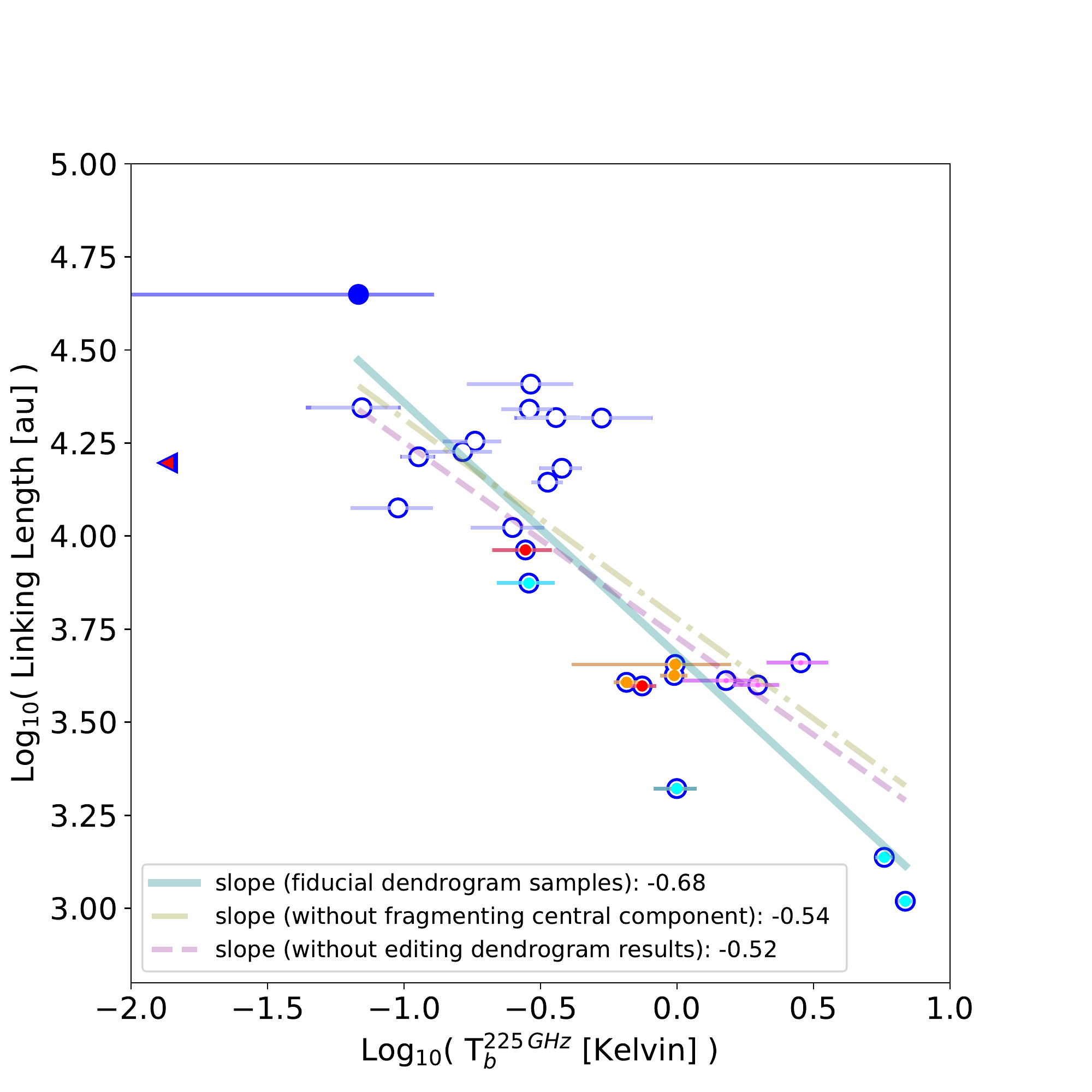} 
  \caption{
    Comparison of the linking lengths among members of the {\it n}-clusters, and the observed continuum brightness temperature ($T_{b}^{\mbox{\tiny 225 GHz}}$) at the centroids of the links. Symbols representing the links of the {\it n}-cluster are color coded in the same way with how the {\it n}-cluster members are color coded in Figure \ref{fig:cluster}. Detections and upper limits are presented by circles and triangles, respectively. Solid line shows the fitted slope of the presented symbols. Dashed-and-dotted and dashed lines show the fittings for the case that the source identified by {\tt dendrogram} at the center is not artificially fragmented into three components, and for the {\tt dendrogram} identification result which is not post processed, respectively.
          }
  \label{fig:fragmentation}
\end{figure}

\section{Discussion}\label{sec:discussion}

\begin{figure}
  \vspace{-1.2cm}
  \hspace{-0.2cm}
  \includegraphics[width=9.7cm]{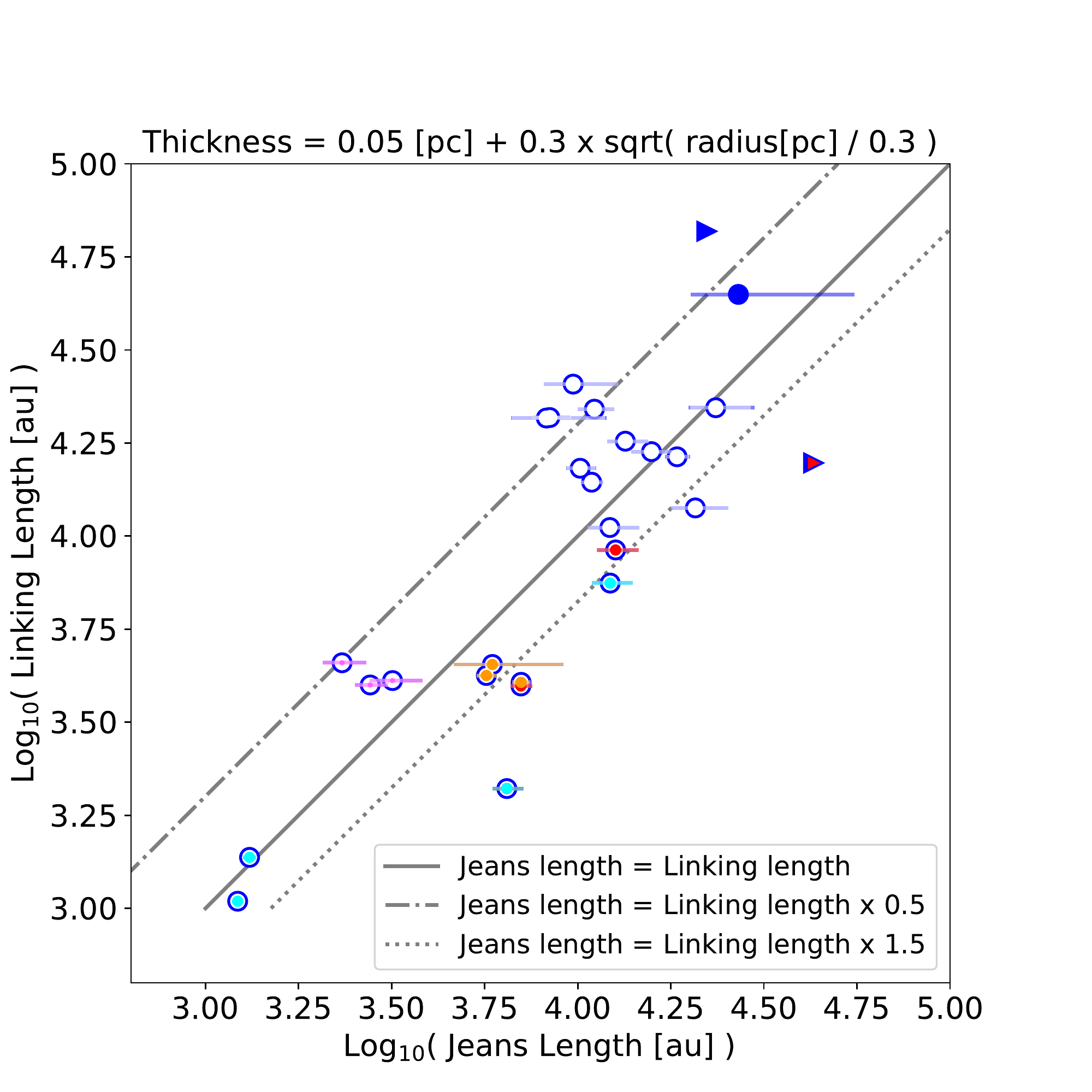}
  \caption{
    Comparison of the linking lengths among members of the {\it n}-clusters, and the derived Jeans lengths. The Jeans lengths were derived based assuming a constant gas temperature of 30 K, and considering that the characteristic thickness $H(r)$ [pc] $=$ 0.05 + 0.3$\times$ $\sqrt{r \mbox{[pc]}/0.3}$, where $r$ is the radius from the centroid of the central 3-cluster (R.A.=18\rah52\ram50\ras.204, Decl.=+00\decd55\decm29\decs.35). Color coding is similar to Figure \ref{fig:fragmentation}. Detections and lower limits are presented by circles and triangles, respectively.
          }
  \label{fig:Jeanslength}
\end{figure}

\subsection{Toomre instability}\label{sub:Toomre}

We discuss the self-gravitational stability/instability of the molecular gas structures by adopting a strategy similar to what was used for analyzing the other OB cluster-forming massive molecular clump, G10.6-0.4 \citep{Liu2017}.
We assume that the dominant motion around G33.92+0.11 is the rotational motion about the center, which is assumed to be at R.A.=18\rah52\ram50\ras.204 and Decl.=$+$00\decd55\decm29\decs.3489 (for more discussion see Section \ref{sub:cluster}).
We assume that the gas structure is flattened, and the centrifugal force of the rotational motion is equal to the gravitational force of the enclosed stellar and gas masses.
We then estimate the enclosed gas mass in radius $r$ based on the observed 225 GHz flux and an assumed gas temperature profile $T(r)$, and considers that the dust emission is optically thin, and that the gas to dust mass ratio is 100.
Finally, we estimate the Toomre Q parameter 
\begin{equation}
Q =  \frac{\kappa c_{s} }{\pi G \Sigma},
\end{equation}
where $\kappa$ is the epicyclic frequency $\sqrt{(2\Omega/r)(d(r^2\Omega)/dr)}$, $\Omega$ is the angular velocity of the rotational motion, $c_{s}$ is the thermal sound speed $\sqrt{k_{b}T /2.3m_{\scriptsize H}}$, $G$ is the gravitational constant, $\Sigma$ is the mass surface density, $k_{b}$ is the Boltzmann constant, and $m_{\scriptsize H}$ is the hydrogen mass.

The gas temperature profile $T(r)$ requires an assumption since we do not know the physical condition at the time epoch when fragmentation occurred.
Motivated by the NH$_{3}$ rotational temperature presented in \citet{Liu2012b}, our fiducial analysis assumes a constant 30\,K temperature across the entire source.
We also present an analysis based on assuming a radial gas temperature profile of
\begin{equation} \label{eq:temperature}
  \begin{split}
    T(r) = 180 [K]\times\sqrt{ 0.0175 / r [pc]} \times \omega(r) + 25 [K]\times(1 - \omega(r) ), \\
    \omega(r) = \exp^{-r [pc] / 0.175},
  \end{split}
\end{equation}
which is motivated by the NH$_{3}$ rotational temperature presented in \citet{Liu2012b}, the CH$_{3}$CN rotational temperature presented in \citet{Liu2015}.
On small scale this temperature profile assumes that the heating is dominated by the centrally embedded OB stars, while on the more extended region the heating may be dominated by intermediate stars and interstellar radiation field.
A weighting factor $\omega(r)$ is assumed to join these two regimes of heating.
We also based on the procedure outlined in \citep{Chen2006} to derive gas temperature from the CH$_{3}$CN J=12-11 lines, which were simultaneously covered by the observations introduced in Section \ref{sec:observation}.
These measurements are in general consistent with the temperature model described by Equation \ref{eq:temperature}.
However, we note that the CH$_{3}$CN line intensities and excitation temperature are azimuthally asymmetric in the inner $\sim$0.1 region, likely due to that it is a complicated cluster-forming environment with multiple heating sources.
A more detailed study of the CH$_{3}$CN lines will be presented in our forthcoming paper (Chen et al. in prep.)
An additional caveat here is that the conversion from NH$_{3}$ rotational temperature is rather empirical \citep[][]{Ho1983}; and while we do not have direct measurements of dust temperature, we also do not know fractionally how much dense gas along a line-of-sight is traced by either CH$_{3}$CN or NH$_{3}$.

Our assumed gas temperature profile, the derived enclosed gas mass, and the evaluated centrifugal velocities and the Toomre Q parameters, are presented in Figure \ref{fig:Toomre}.
G33.92+0.11 already contains an ultra compact (UC) H\textsc{ii} region \citep{Liu2012b}.
Presently, we do not know exactly where the embedded OB stars are located, and what are their exact masses.
Therefore, in our analysis we provide a comparison of the results based on the two rather extreme assumptions: (1) there is no star embedded at the center of the system, and (2) there is a total of 30 $M_{\odot}$ of stellar mass embedded at the center of the system.
We argue that the reality is likely in between these two rather extreme cases.

Based on any of the aforementioned assumptions, the system appears Toomre unstable (i.e., Q$<$1) on the resolved spatial scales.
This is consistent with the spatially resolved hierarchical fragmentation from the continuum image (Figure \ref{fig:almacycle4}).
Including both the effects of the embedded stellar mass and the higher gas temperature may make the value of Q close to 1.0 on $\sim$1000 AU scales.
However, $\Sigma$ may be underestimated on such small spatial scales since some embedded denser structures may be optically thick.
We note that the directly observed gas line-of-sight velocities are considerably smaller than any of the velocity model presented in Figure \ref{fig:Toomre} \citep[c.f.,][]{Liu2015}, which we interpret as an effect of projection.
The fragmentation of core A1 may be suppressed once a (or some) centrally embedded star(s) have achieved higher stellar mass(es).

We remark that whether or not there can form centrifugally supported gas structures is related less directly to the absolute spatial scales, but is more related to the initial fraction of energy that is in the form of rotational motion.
We here take the case of our target source, G33.92+0.11 as an example.
\citet{Liu2015} reports that the enclosed gas mass within a $\sim$5 pc radius is $\sim$10$^{5}$ $M_{\odot}$.
Assuming that the cloud is initially, marginally virialized, then the characteristic gas velocity at the $\sim$5 pc effective radius is $\sim$5 km\,s$^{-1}$ (i.e., the projected component along the line-of-sight $\sim$3 km\,s$^{-1}$).
Assuming that the $\sim$1\% of the gas kinetic energy is in the form of rotational motion, then the averaged rotation motion at $\sim$5 pc radius is on the order of $\sim$0.5 km\,$^{-1}$.
Assuming that the global collapse conserves specific angular momentum, then the velocity of the rotational motions will be amplified to $\sim$5 km\,s$^{-1}$ at a $\sim$0.5 pc radius, which is just enough to support against the gravity of the $\sim$$10^{3}$ $M_{\odot}$ enclosed gas and stellar masses within that radius.
While we do not know how we should assume the energy ratio, it appears to us that an 1\%-2\% of energy in rotational motions is an reasonable assumption when comparing with the galactic wide surveys \citep[e.g.,][]{Braine2018}.
Strong magnetic field with a very specific geometric configuration can propagate out the specific angular momentum.
However, the strengths and geometric configurations of magnetic field in molecular clouds is yet a matter of debates.
In our previous work, we found that the collapse on 0.5-1.5 pc scales may indeed conserve specific angular momentum \citep{Liu2010}.

\begin{figure}
  \vspace{-7.5cm}
  \hspace{-0.3cm}
  \includegraphics[width=9.5cm]{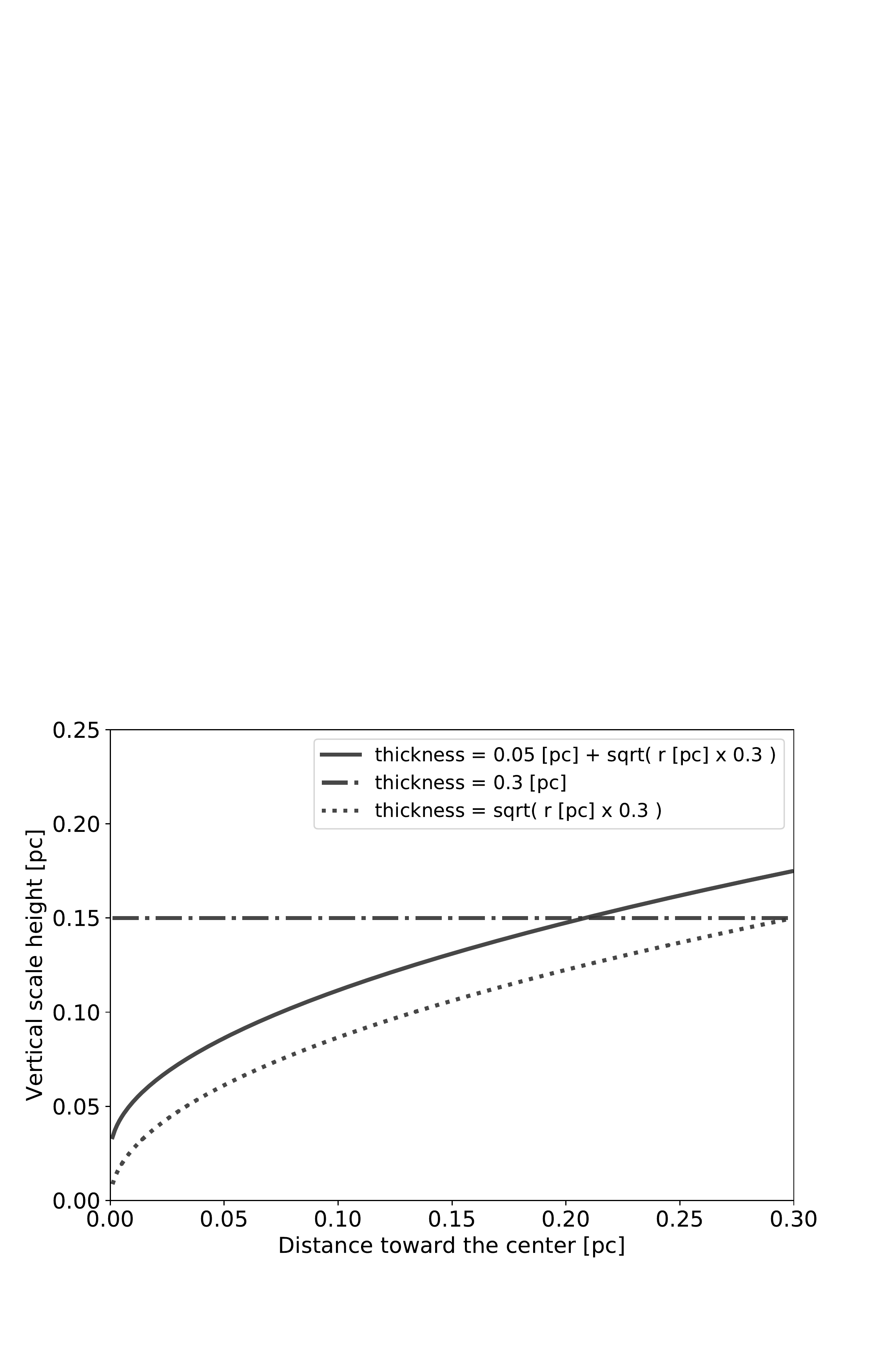}
  \vspace{-1.5cm}
  \caption{The assumptions of the vertical scale height (i.e., half of the characteristic thickness, see Section \ref{sub:cluster}.}
  \label{fig:thickness}
\end{figure}

\subsection{Identifying candidates of young stellar objects}\label{sub:dendrogram} 

To identify Class 0/I candidates from the primary beam uncorrected, Briggs Robust = 0 weighted continuum image, we first used the {\tt dendrogram} algorithm \citep{Rosolowsky2008} to yield a complete list of independent sources (i.e., the {\it leafs} in the terminology of {\tt dendrogram}), and then systematically pick out the relatively compact sources from the list based on our own empirically defined criteria.
When running {\tt dendrogram}, we set the minimum pixel value to be two times our RMS noise level, and set the minimum delta pixel value to be five times our RMS noise level.
Intensity fluctuations smaller than the minimum delta pixel value are not regarded as real structures.
We note that the expected number of spuriously identified sources can be estimated as a product of the total number of independent measurements and the probability that the noise is higher than the minimum delta pixel value.
For our present case, the primary beam FHWM is approximately resolved by $7\times10^{4}$ independent synthesized beams.
By setting the 5-$\sigma$ noise level as the minimum delta pixel value, assuming Gaussian noise distribution, the expected number of spuriously identified sources is 0.04, while it is $\sim$5 and $\sim$200 if by setting the 4-$\sigma$ and 3-$\sigma$ noise levels as the minimum delta pixel value.

We required the identified structures to have areas more extended than 25 pixel$^{2}$, which is approximately the synthesized beam area of our image in terms of 1-$\sigma$ Gaussian beam width.
Running {\tt dendrogram} using these input parameters yielded an initial catalogue of 33 identified sources within our primary beam FWHM ($\sim$28$''$).
In this catalogue, there are four identified sources associated with A1, where the central one presents an irregular morphology likely because it in fact consists of multiple internal sources.
The southern most Class 0/I candidate we identified may be a binary source.
Presently we cannot be very sure since it is located too close to the edge of our primary beam FWHM to be imaged well.
Nevertheless, this will not make a significant impact to our following discussion.

We selected the sources with high contrast from the initial catalogue, which are more likely to be true Class\,0/I objects, by requiring them to further fulfill at least one of the following three conditions:
\begin{itemize}
  \item[1.] The radius derived from fitting an ellipse is smaller than 8 pixels (0\farcs08, $\sim$570 AU), and the peak intensity is higher than five times the RMS noise. \\\vspace{-0.65cm}
  \item[2.] The ratio in between the highest and the lowest intensity within the identified structure is higher than 12. \\\vspace{-0.65cm}
  \item[3.] The source is located outside of the 5$''$ radius from the center, such that it is less subject to confusion with ambient dense gas structures. \\\vspace{-0.5cm}
\end{itemize}
And finally, we artificially fragment the centralized source embedded at the center of A1 to be three sources, to meet our visual impression.
In the end we obtain a catalogue with a total of 28 Class 0/I candidates, which are presented in Figure \ref{fig:cluster} (the original {\tt dendrogram} output is provided in Appendix \ref{appendix:dendrogram}).
Our edits to the identified source catalogue is subjective to some extent.
Our discussion will be mainly based on the edited catalogue, which meets our visual impression better. 
On the other hand, we had also analyzed the initial, unedited catalogue to demonstrate that our edits did not largely bias our conclusion.
These subjective edits are mainly due to the yet limited spatial resolution of our image.
Once the angular resolution is improved by future observations (e.g., to $\sim$200 AU), the first two criteria may be replaced by directly checking certain physical properties of the circumstellar envelope (e.g., radial column density profile), and the third criterion may be relaxed due to the suppressed confusion limit.
In the Briggs Robust = -2 weighted continuum image which has slightly higher angular resolution, the Class 0/I candidates appear more visually distinguishable from the ambient gas structures, although its higher RMS noise also confuses the  {\tt dendrogram} identification.

\begin{figure}
  \vspace{-1cm}
  \hspace{-0.8cm}
  \begin{tabular}{ p{8.7cm} }
    \includegraphics[width=9.7cm]{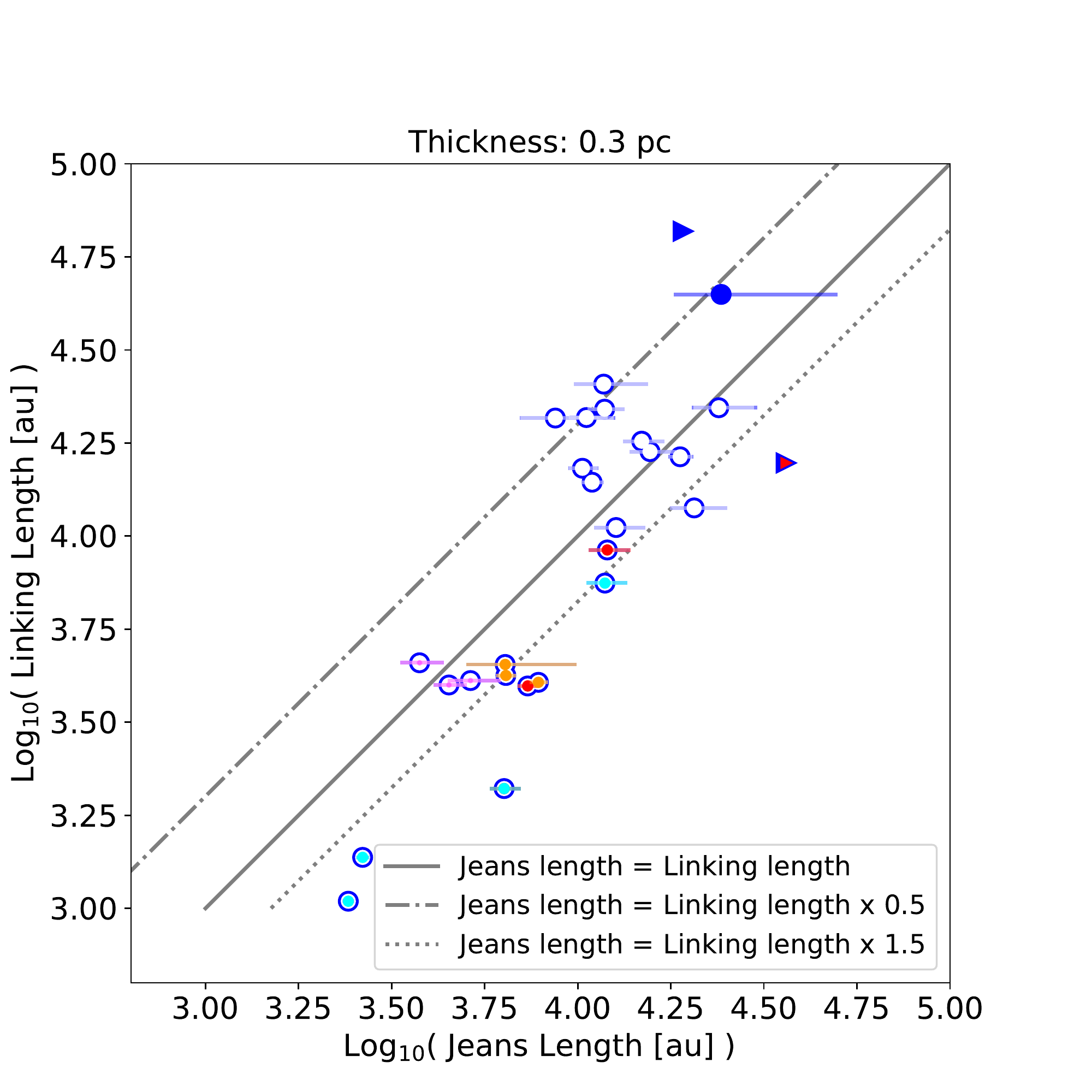} \\
    \vspace{-1.5cm}\includegraphics[width=9.7cm]{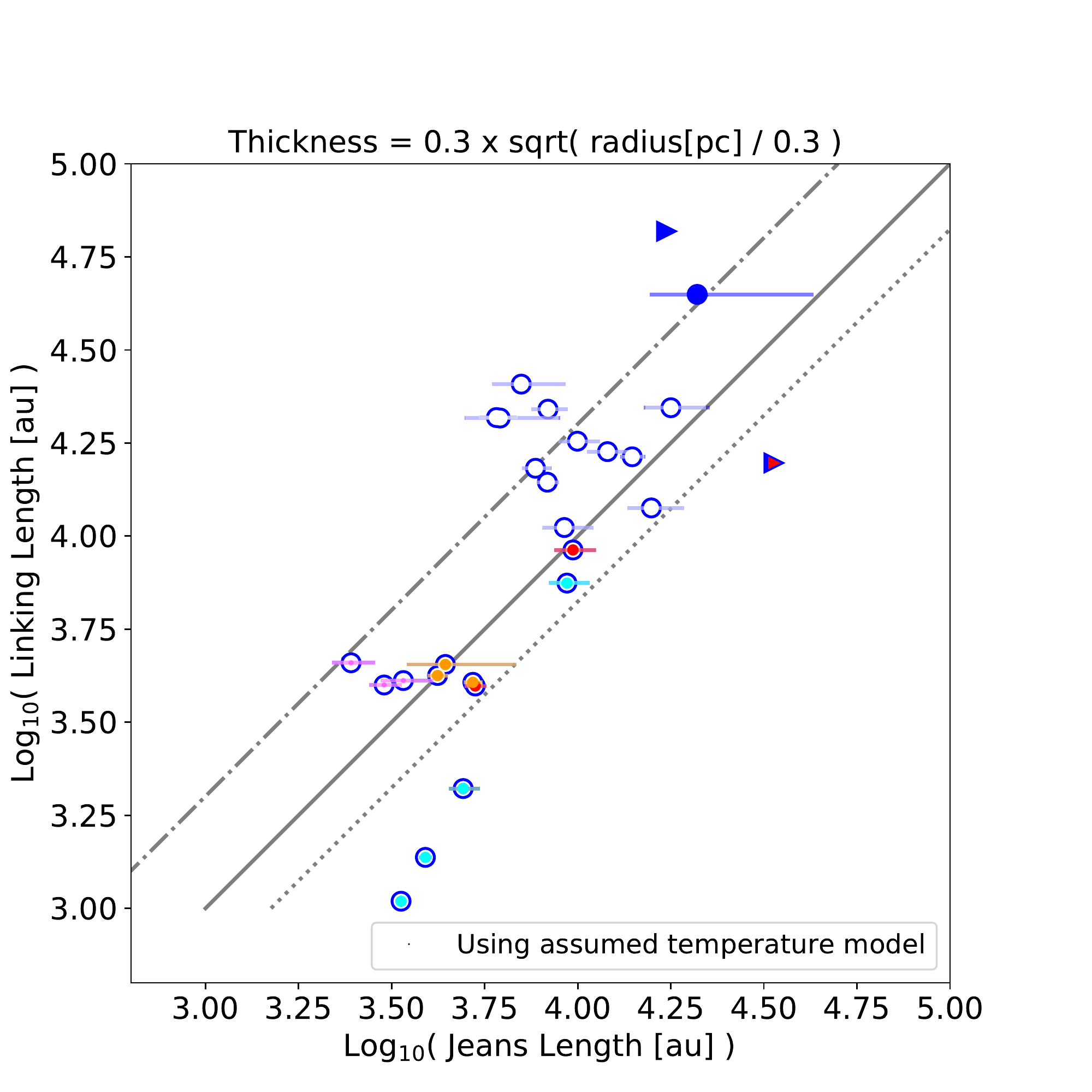} \\
  \end{tabular}
  \caption{
    Similar to Figure \ref{fig:Jeanslength}. Top panel shows the Jeans lengths derived based on an assumption of a constant characteristic thickness 0.3 pc and a constant temperature 30 K; bottom panel shows the Jeans lengths derived based on assuming the characteristic thickness $H(r)$ [pc] $=$ 0.3$\times$ $\sqrt{r \mbox{[pc]}/0.3}$ and use the assumption of the spatially varying temperature profile presented in Figure \ref{fig:Toomre}. Color coding is similar to Figure \ref{fig:fragmentation}.
          }
  \label{fig:Jeanslength2}
\end{figure}

\subsection{Clustering of candidates of young stellar objects}\label{sub:cluster}

We have performed a mean surface density of companions (MSDC) analysis to our source list. 
The MSDC is often used to study the spatial distributions of stars and other 
sources (e.g. continuum peaks) in 2-dimensional configurations and investigate 
the scale regimes at which a system moves from a regular distribution of pairs to 
clusters \citep[e.g.][]{gomez93,bate98,roman10,tafalla15,Palau2018}. 
We followed the prescription of \citet{simon97}, also described in detail in 
 \citep{Palau2018}, and we summarize it here briefly: we estimated angular 
 separations, $\theta$ between each other source within the list and 
 organized the result in annular bins of 0.2 dex within 
 $-4.5<\log{\theta}<-1.5$. 
 We computed the MSDC, $\Sigma _\theta$ as the number of elements in each annuli, $N_p(\theta)$ divided by the area and normalizing by the total number of separations. 
 We also estimated the two-point correlation function, TPCF, that compares $\Sigma _\theta$ with a random distribution of points in the same area, $A_m$ (in our case a circle of 28 arcsec in diameter)  as $W_\theta=(N_p/N_r)-1$. The random distribution was calculated from $10^4$ distributions constructed with a Monte Carlo routine.
This way, we can recalculate $\Sigma_\theta=(N_p/A_m)(1+W_\theta)$ and we can compare both versions of $\Sigma _\theta$ directly; when the points are similar, we can assure that those points are reliable, while a separation of the two estimates by more than the typical uncertainty can be due to edge effects in our finite distribution and for those points the MSDC function is less reliable. 
The MSDC analysis result is shown in Figure \ref{fig:MSDC}. 
We see that the points are clearly linearly anti-correlated within  $-3.75<\log{\theta}<-3.3$ (0.022 to 0.062 pc, or roughly 4500 to 12800 AU. 
This linear regime is consistent with a regular distribution of the sources along the filamentary structure at the distinct scales, and the very steep slope of -3.6 may be indicative of the spiral morphology of the filaments (regular elongated filaments should have a slope close to -1). 
Then the function shows a clear bump structure within $-3.3<\log{\theta}<-2.65$ (0.062 to 0.28 pc or 12800 to 57100 AU) with a peak near $\log{\theta}=-3.0$ (25500 AU). 
This bump would be consistent with a clustered distribution which may reflect the aggregation of the 
peaks as a whole. 
The MSDC looses reliability beyond 0.3 pc due to the limited size of our map. Also interesting is the fact that we see agreement between the two estimates of the MSDC down to $\log{\theta}=-4.4$, close to the 1000 AU regime, indicating how we are resolving fragmentation down to those scales.
In general, the MSDC analysis has shown that the identified Class 0/I candidates are likely grouped into certain patterns and well distributed along the filamentary structure instead of being randomly distributed.

To better visualize the clustering of the Class 0/I candidates, and also for the convenience of our discussion, we define a {\it hierarchical n-cluster} as a set of Class 0/I candidates which fulfill the follow criteria:

\begin{itemize}
  \item[(i)] The number of Class\,0/I candidates in this set is an integer $n$. \\\vspace{-0.65cm}
  \item[(ii)] The nearest Class\,0/I candidate of any member in this set is also a member of this set. \\\vspace{-0.65cm}
  \item[(iii)] We define {\it links} as the fewest required line segments, which are also the shortest possible ones, to allow visiting all cluster members by following the links. A set including all links forms the {\tt minimum spanning tree} of the cluster. For example, a 3-cluster has two links. When two members of a 3-cluster are the nearest neighbors of each other, then the links are the line segment connecting these two members, and the shorter one of the two line segments which connect individual of these two members with the remaining one. We require the longest link of a n-cluster to be shorter than the smallest separation from its members to the non-member Class 0/I candidates by a factor of a certain separation threshold $\xi$. \\\vspace{-0.65cm}
\end{itemize}

By introducing this concept, we are essentially identifying clusters based on the relative spatial separations of sources instead of the absolute spatial separations, which is useful for the studies of the hierarchical self-gravitational fragmentation when the systems are resolved with broad ranges of density or temperature.
In this paper, we chose $\xi$ to be 1.7.
In the case that the gas temperature is uniform, this $\xi$ value signifies the variations of Jeans length due to a factor $\sim$3 changes of gas density when forming the dense gas fragments.
The value of $\xi$ is not important for any physics we discuss within the present manuscript.
Based on this concept, we identified four 2-cluster, two 3-cluster, one 4-cluster, one 6-cluster, one 25-cluster, and all 28 identified Class\,0/I candidates as a 28-cluster.
In Figure \ref{fig:cluster}, we color coded the identified Class\,0/I candidates (by {\tt dendrogram}) according to how they are associated with those n-clusters.
For example, A1 (Figure \ref{fig:almacycle4}) is associated with a 6-cluster, where the central three members of this 6-cluster is also a 3-cluster.
We tentatively adopt the center of the system as the averaged projected position of the members of this 3-cluster, R.A.=18\rah52\ram50\ras.204 and Decl.=$+$00\decd55\decm29\decs.3489.
We think this assumption of center is reasonable, since for all observed spatial scales the density distribution about the center will not be dramatically lopsided.
It is also spatially very close to the dominant source(s) of ionizing photons according to the geometry of the UC H\textsc{ii} region \citep{Liu2012b}.
Similarly, A2 (Figure \ref{fig:almacycle4}) is associated with a 4-cluster.
The Class\,0/I candidates in the inner $\sim$0.3 pc radius is a 25-cluster, which is also the central part of the largest identified (28-)cluster.
The largest cluster may have more members, which are located outside of the primary beam FWHM of our ALMA observations.

\subsection{Spatial separations of young stellar objects}\label{sub:separation}

The {\it links} we defined in Section \ref{sub:cluster} naturally pass through the elongated or arm-like dense gas structures which host the Class\,0/I candidates.
If the Class\,0/I candidates are indeed representative of the dense gas fragments forming out of the elongated or arm-like dense gas structures, then the lengths of the links, which we refer to as {\it linking lengths}, are expected to be inversely correlated with the square root of the gas density in these elongated or arm-like dense gas structures.
We hypothesize that the physical properties measured at the center of the links may be representative of the physical condition when fragmentation occurred.

Figure \ref{fig:fragmentation} shows a comparison of the linking lengths with the dust brightness temperature $T_{b}^{\mbox{\scriptsize 225 GHz}}$ measured at the center of the links.
For each link which connects two Class 0/I candidates, we nominally took the standard deviation of $T_{b}^{\mbox{\scriptsize 225 GHz}}$ measured along the inner 50\% portion of the link as the error. 
We present $T_{b}^{\mbox{\scriptsize 225 GHz}}$ plus and minus one error for links of which $T_{b}^{\mbox{\scriptsize 225 GHz}}$ is above our RMS noise level, otherwise present one error as upper limits.
Interestingly, the linking lengths $\ell$ are indeed approximately inversely proportional to $\sqrt{T_{b}^{\mbox{\scriptsize 225 GHz}}}$, which is a tracer of gas column density $\Sigma$.
We are not convinced that the relation in between $\sqrt{T_{b}^{\mbox{\scriptsize 225 GHz}}}$ and $\Sigma$ has a dependence on the number ($n$) of cluster members, although the numbers of identified $n$-clusters remain small for any $n$.
A linear regression of $\log{(\sqrt{T_{b}^{\mbox{\scriptsize 225 GHz}}})}$ and $\log{(\ell)}$ (without including upper limits) for the Class\,0/I candidates in our final catalogue (Section \ref{sub:dendrogram}) gives a slope of -0.68.
The links in between the members of the central 3-cluster appear shorter than what are indicated by the regression line.
The same analysis for the Class\,0/I candidates in the catalogue that we did not artificially fragment the central source, and for the catalogue directly given by {\tt deodrogram} without any further edits, gives slopes of -0.54 and -0.52, respectively.

To explain Figure \ref{fig:fragmentation}, we convert $T_{b}^{\mbox{\scriptsize 225 GHz}}$ to gas volume density $\rho$ based on the assumed radial characteristic thickness profile $H(r)$, and assuming $\kappa_{\mbox{\scriptsize 225 GHz}}=$0.6 cm$^{2}$\,g$^{-1}$ and gas to dust mass ratio 100.
We then further convert $\rho$ to Jeans lengths based on the assumed radial temperature profile $T(r)$.
We note that $H(r)$ cannot be directly measured.
Our assumption of $H(r)$ is motivated by observations of the similar but edge-on system, G10.6-0.4 \citep{Liu2017}.
Figure \ref{fig:Jeanslength} shows our fiducial case which assumes a constant temperature 30 K, and the characteristic thickness at radius $r$
\[
H(r) [pc] = 0.05 + 0.3\times\sqrt{r [pc] / 0.3}.
\]
Geometrically this is a flattened system, for which the inner part is flatter than the outer part (Figure \ref{fig:thickness}).
The derived Jeans lengths based on these assumptions are consistent with the linking lengths $\ell$.
The flatter inner region of the system may be explained by the compression by gravitational force.
When we view the central 3-cluster as a single entity (i.e., without artificially fragmenting the central source of the {\tt dendrogram} catalogue), then the effects of the spatially varying scale height become less obvious, 
and therefore $\ell$ appears better correlated with $\sqrt{T_{b}^{\mbox{\scriptsize 225 GHz}}}$.

This assumption of the thickness of gas structure is likely yet oversimplified.
This problem is general in this research topic, and presently we do not find a vastly better way of handling it.
The gas structure around the other massive core, A2, may also be compressed in the vertical direction by (self-)gravity.
This is not considered in our simplified assumption of thickness.
As a consequence, the evaluated Jeans lengths for the 4-cluster associated with A2 appear slightly larger than the observed $\ell$ (Figure \ref{fig:Jeanslength}).

To provide a sense about how the uncertainties in thickness and gas temperature is propagated to the evaluated Jeans lengths, in Figure \ref{fig:Jeanslength2}, we compare the measured $\ell$ with the Jeans lengths which were estimated based on the assumption of constant 30 K temperature and 0.3 pc thickness across the entire system; and the Jeans lengths which were estimated based on the assumption of the radially varying temperature profile as presented in the top left panel of Figure \ref{fig:Toomre} and the characteristic thickness profile $H(r) [pc] = 0.3\times\sqrt{r [pc] / 0.3}$.
These comparisons show that the Jeans lengths around the central 3-cluster may be too high as compared to the measured $\ell$ if the thickness of the system is not reduced at the center.
The constant $\sim$0.3 pc thickness may still be reasonable for the rest part of the system.
In addition, the Jeans lengths around the central 3-cluster may also be too high if this region is already significantly heated by the central OB cluster prior to the formation of this 3-cluster.
However, we cannot rule out that there are Class\,0/I objects forming at wider spatial separations but migrate toward the center afterwards to form the central 3-cluster.

In general, we think that the formation of the Class\,0/I candidates in G33.92+0.11 is consistent with self-gravitational fragmentation of a geometrically flattened, Toomre unstable rotating accretion flow.
It is not yet clear how (micro-)turbulence \citep[e.g.,][]{White1977} can support gas structures from fragmentation.
However, based on the analysis presented in this section, we consider that for this particular target source, it may be fair to not include the effect of (micro-)turbulence in any way when estimating the Toomre Q parameter (c.f., Section \ref{sub:Toomre}).
There were similar suggestions made by \citet{Palau2015,Palau2018} based on the analysis of the fragmentation of the OMC-1S region, and by the Northern Extended Millimeter Array CORE survey \citep{Beuther2018}.
This hypothesis is also indirectly supported by the observations of subsonic molecular gas linewidths in some high-mass star-forming regions \citep[e.g.,][and references therein]{Monsch2018,Sokolov2018}.
We also refer to \citet{Pokhrel2018} for a related suggestion.
If the fragmentation process is indeed governed mainly by thermal pressure and self-gravity, then the numbers of the identified $n$-clusters may be regarded as footprints of the three dimensional gas volume density distribution during when the fragmentation occurred.
A caveat here is that Jeans length is only a prescription for describing the fragmentation of a spatially uniform, infinitely extended, matter distribution, starting everywhere at zero velocity.
For a dynamically collapsing system with finite size, the characteristic spatial separation of fragments may be still comparable to the Jeans lengths \citep[e.g.,][]{Larson1985}.
More insight may be provided from numerical hydrodynamics simulations, which is beyond the scope of the present paper.
On the other hand, even based on the assumption that the Class\,0/I candidates are forming out of self-gravitational fragmentation, our fiducial assumption of $H(r)$ still cannot be treated as a measurement of $H(r)$.

With the present experience of resolving A1 and A2 into multiple components \citep[Figure \ref{fig:almacycle4};][]{Liu2012b,Liu2015}, we are concerned that we are not yet resolving the smallest gas fragments to form individual high- and low-mass stars.
Therefore, we defer the studies of Jeans mass to future, higher angular resolution observations.

\section{Conclusion}\label{sec:conclusion}

We have performed $\sim$1000 AU spatial resolution observations at $\sim$225 GHz with ALMA towards the flattened and rotating OB cluster-forming clump, G33.92+0.11.
The target source is likely in a face-on projection, which is ideal for the studies of the morphology of dense gas.
We found that dense gas in this region is Toomre unstable, forming the $\sim$0.3 pc scales arm-like or elongated structures, which are hierarchically embedded with smaller internal sub-structures.
The spatial separations of the candidates of Class 0/I YSOs we identified from this source can be explained with Jeans lengths, which were derived based on rather simple assumptions.
Our present interpretation is in concert with the previously observed fact that the virial parameters on $\sim$1 pc scale and on the spatial scales of individual dense cores appear small, which indicates that the hierarchically forming dense gas structures are supported by specific angular momentum in the direction approximately along our line-of-sight.

\acknowledgments %
{\it We thank our referee for the very useful report.}
This paper makes use of the following ALMA data: ADS/JAO.ALMA \#2012.1.00387.S,  \#2016.1.00362.S. ALMA is a partnership of ESO (representing its member states), NSF (USA) and NINS (Japan), together with NRC (Canada), MOST and ASIAA (Taiwan), and KASI (Republic of Korea), in cooperation with the Republic of Chile. The Joint ALMA Observatory is operated by ESO, AUI/NRAO and NAOJ.
PH acknowledges support from the Ministry of Science and Technology of Taiwan with Grant MOST 105-2112-M-001-025-MY3.
I.J.-S. acknowledges financial support from the STFC through an Ernest Rutherford Fellowship (proposal number ST/L004801).

\facility{ALMA}
\software{CASA \citep{McMullin2007}, Numpy \citep{VanDerWalt2011}, APLpy \citep{Robitaille2012}, astrodendro (https://github.com/dendrograms/astrodendro/tree/stable)}

\bibliographystyle{yahapj}
\bibliography{references}

\appendix

\section{Dendrogram analysis}\label{appendix:dendrogram}
\begin{figure}
\vspace{-10cm}
\hspace{-1cm}
\includegraphics[width=35cm]{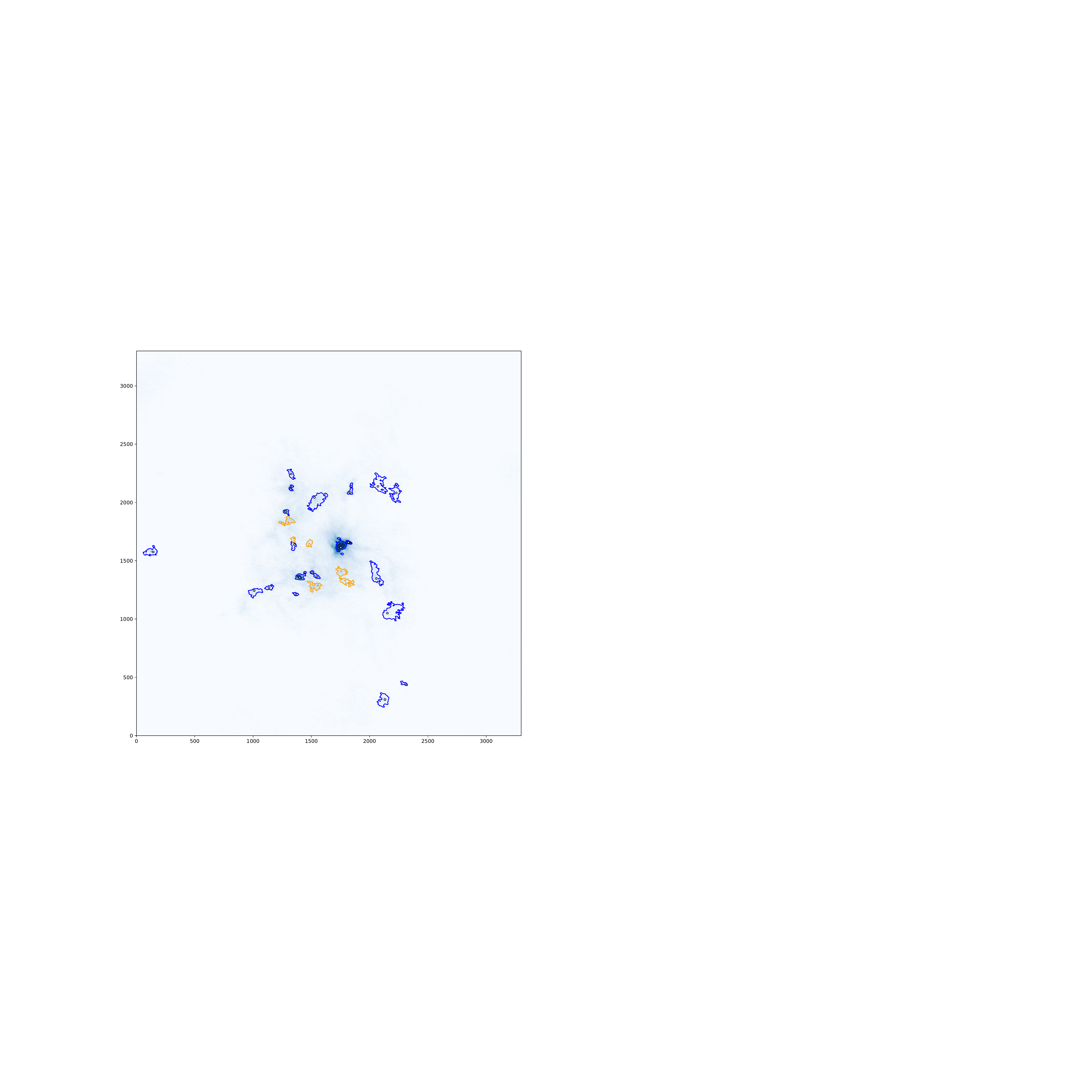}
\vspace{-10cm}
\caption{
  Color image is the same with the one used for Figure \ref{fig:cluster}. Contours are the original output from {\tt dendrogram} for identified regions: blue and orange colors indicate those which were preserved/removed after implementing our additional criteria (for details see Section \ref{sub:dendrogram}). The units of the horizontal and vertical axes are pixels, while we require the area of individual identified sources to be wider than 25 pixel$^{2}$.
        }
   \label{fig:dendrogram}
\end{figure}

Figure \ref{fig:dendrogram} shows the original output from {\tt dendrogram} for identified regions.

\section{Velocity channel maps of the SiO 5-4 line}\label{appendix:sio}

We present the velocity channel maps of the SiO 5-4 line in Figure \ref{fig:sio}.
Our visual impression is that all of the SiO 5-4 emission features (e.g., collimated jets, bow shock features) are associated with our identified Class\,0/I candidates; and conversely, most of our identified Class\,0/I candidates are associated with SiO 5-4 emission features.
For example, the SiO emission associated with the source C (Figure \ref{fig:almacycle4}) can be seen from the velocity channels of 104.7-108.9 km\,s$^{-1}$.
The SiO emission associated with the two southern most sources can be seen from the velocity channels of 103.3-111.7 km\,s$^{-1}$.
Systematically (and non-manually) associating the SiO emission features to the identified Class 0/I candidates, is however, not necessarily trivial.
For example, the three Class\,0/I candidates southeast of the A2 core are likely the powering sources of the extended bow shock like features southwest of them, which can be seen from the velocity channels of 104.7-110.3 km\,$^{-1}$.
However, these bow shock like features are not directly connected with the three Class\,0/I candidates by collimated SiO jets.
This can be either because that these three Class\,0/I candidates were undergoing episodic accretion and jet knot eruptions, or may because that the SiO 5-4 line has too high excitation temperature such that it does not trace the lower temperature parts of the jets/outflows.
Such uncertainty is the reason why we refer to the {\tt dendrogram} identified sources as Class\,0/I candidates, instead of confirmed Class 0/I YSOs.
This can be testified by future, dedicated observations of molecular line tracers for outflows/jets and circumstellar disks/envelopes.

\figsetstart
\figsetnum{1}
\figsettitle{Velocity channel maps of SiO 5-4}
\figsetgrpstart
\figsetgrpnum{figurenumber.1}
\figsetgrptitle{Velocity channel maps set 1.}
\figsetplot{sio_channel_g33p92_1.pdf}
\figsetgrpnote{Velocity channel maps of SiO 5-4.}
\figsetgrptitle{Velocity channel maps set 2.}
\figsetplot{sio_channel_g33p92_2.pdf}
\figsetgrpnote{Velocity channel maps of SiO 5-4.}
\figsetgrptitle{Velocity channel maps set 3.}
\figsetplot{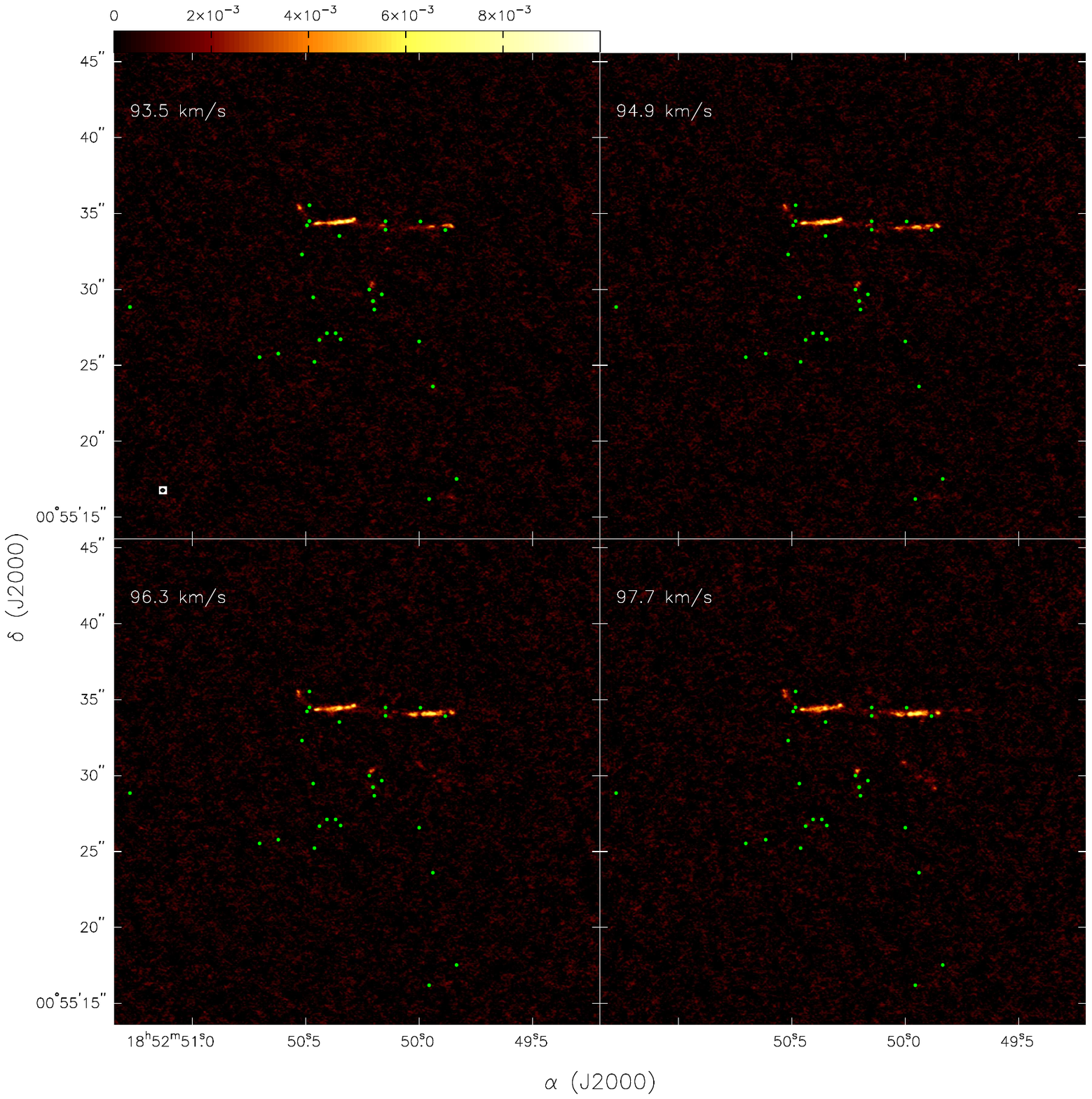}
\figsetgrpnote{Velocity channel maps of SiO 5-4.}
\figsetgrptitle{Velocity channel maps set 4.}
\figsetplot{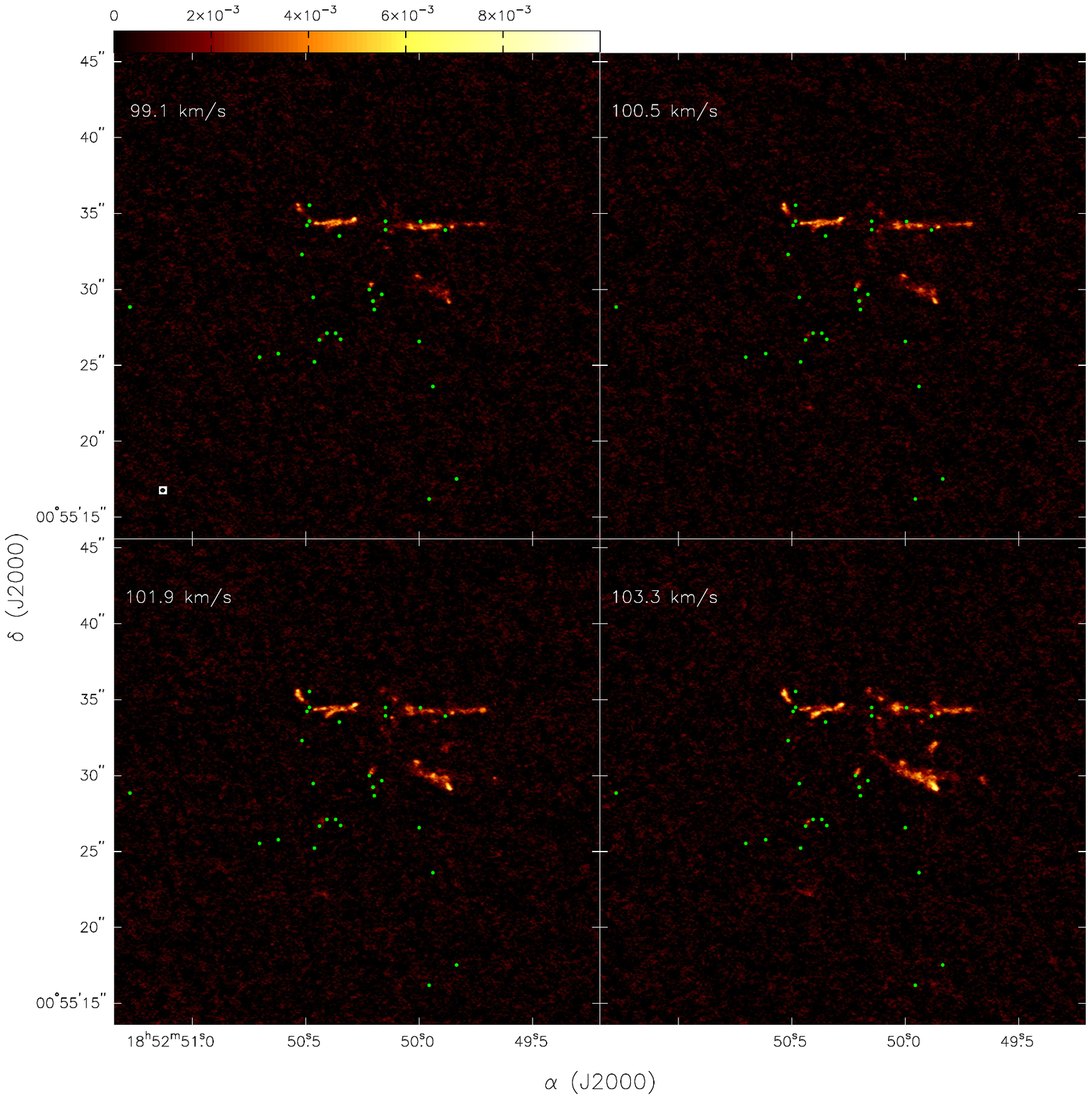}
\figsetgrpnote{Velocity channel maps of SiO 5-4.}
\figsetgrptitle{Velocity channel maps set 5.}
\figsetplot{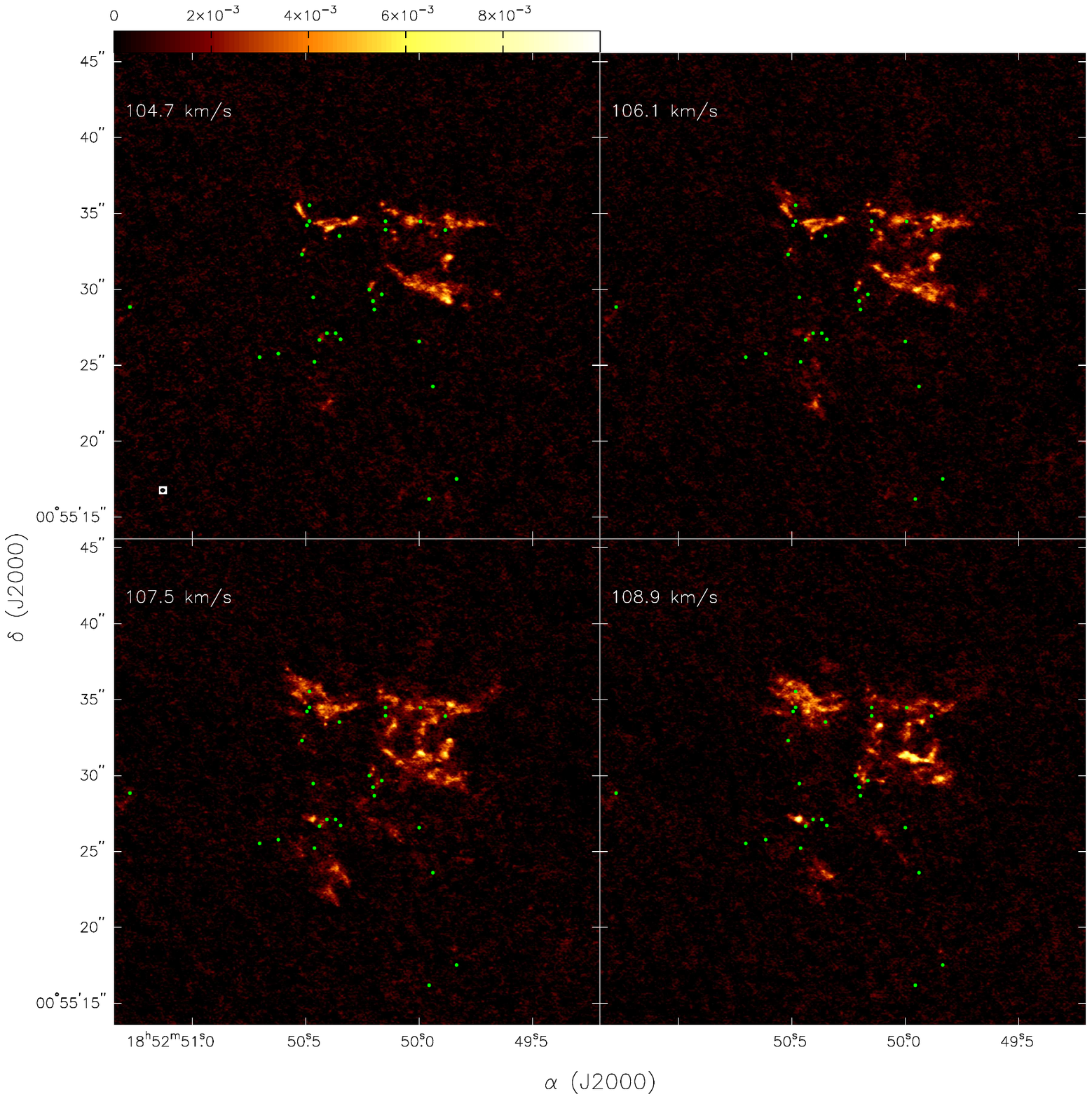}
\figsetgrpnote{Velocity channel maps of SiO 5-4.}
\figsetgrptitle{Velocity channel maps set 6.}
\figsetplot{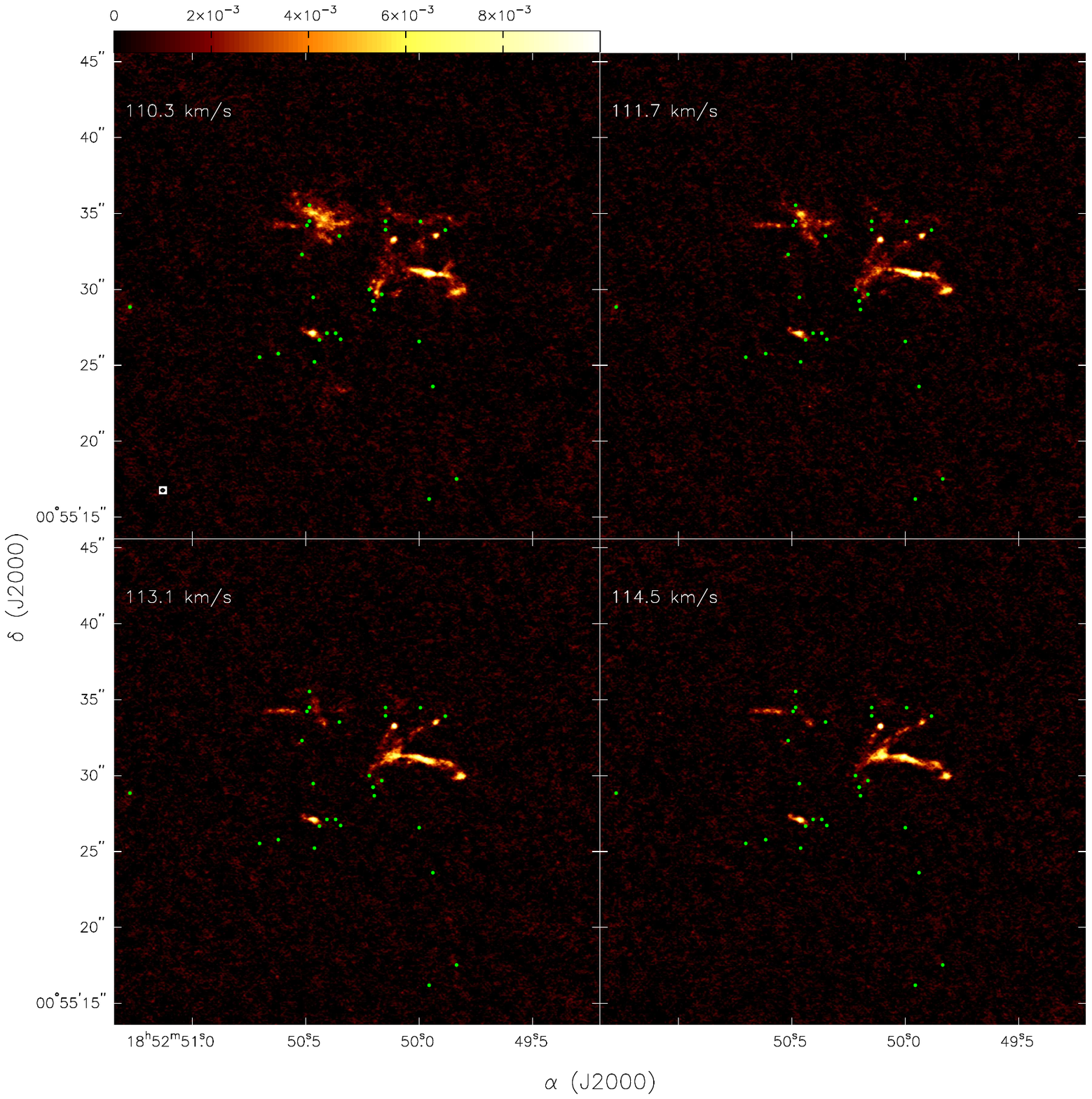}
\figsetgrpnote{Velocity channel maps of SiO 5-4.}
\figsetgrptitle{Velocity channel maps set 7.}
\figsetplot{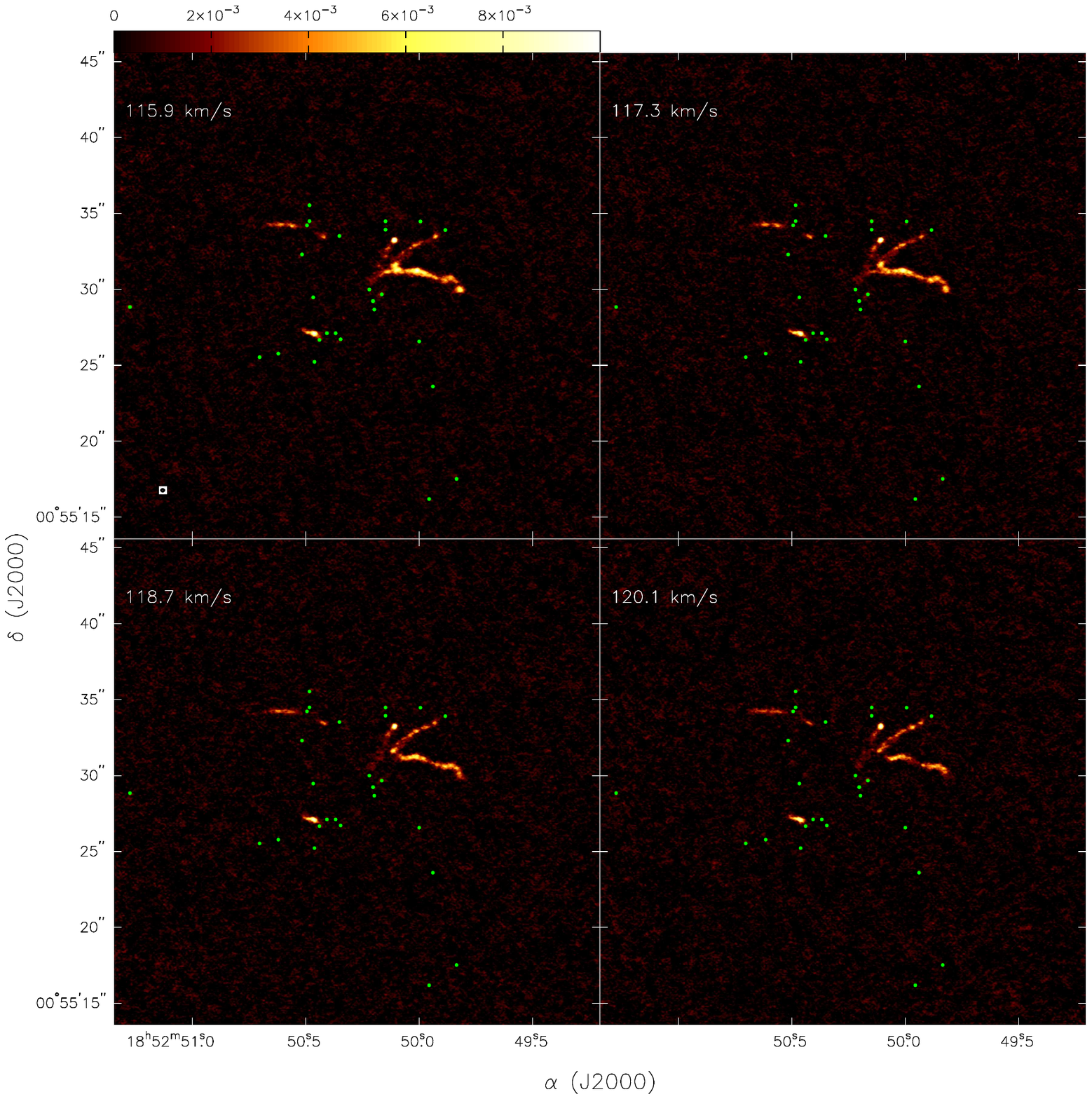}
\figsetgrpnote{Velocity channel maps of SiO 5-4.}
\figsetgrptitle{Velocity channel maps set 8.}
\figsetplot{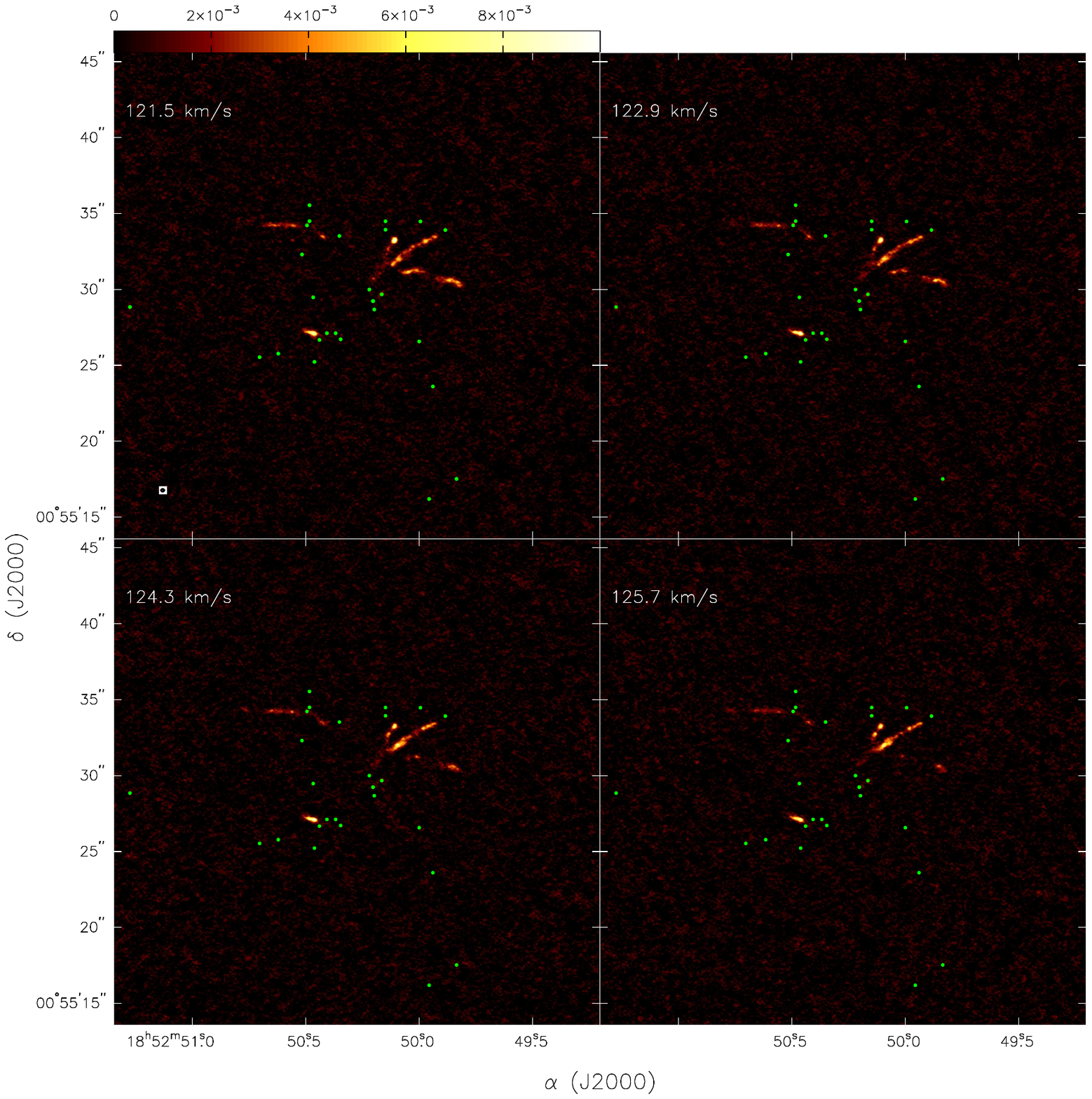}
\figsetgrpnote{Velocity channel maps of SiO 5-4.}
\figsetgrptitle{Velocity channel maps set 9.}
\figsetplot{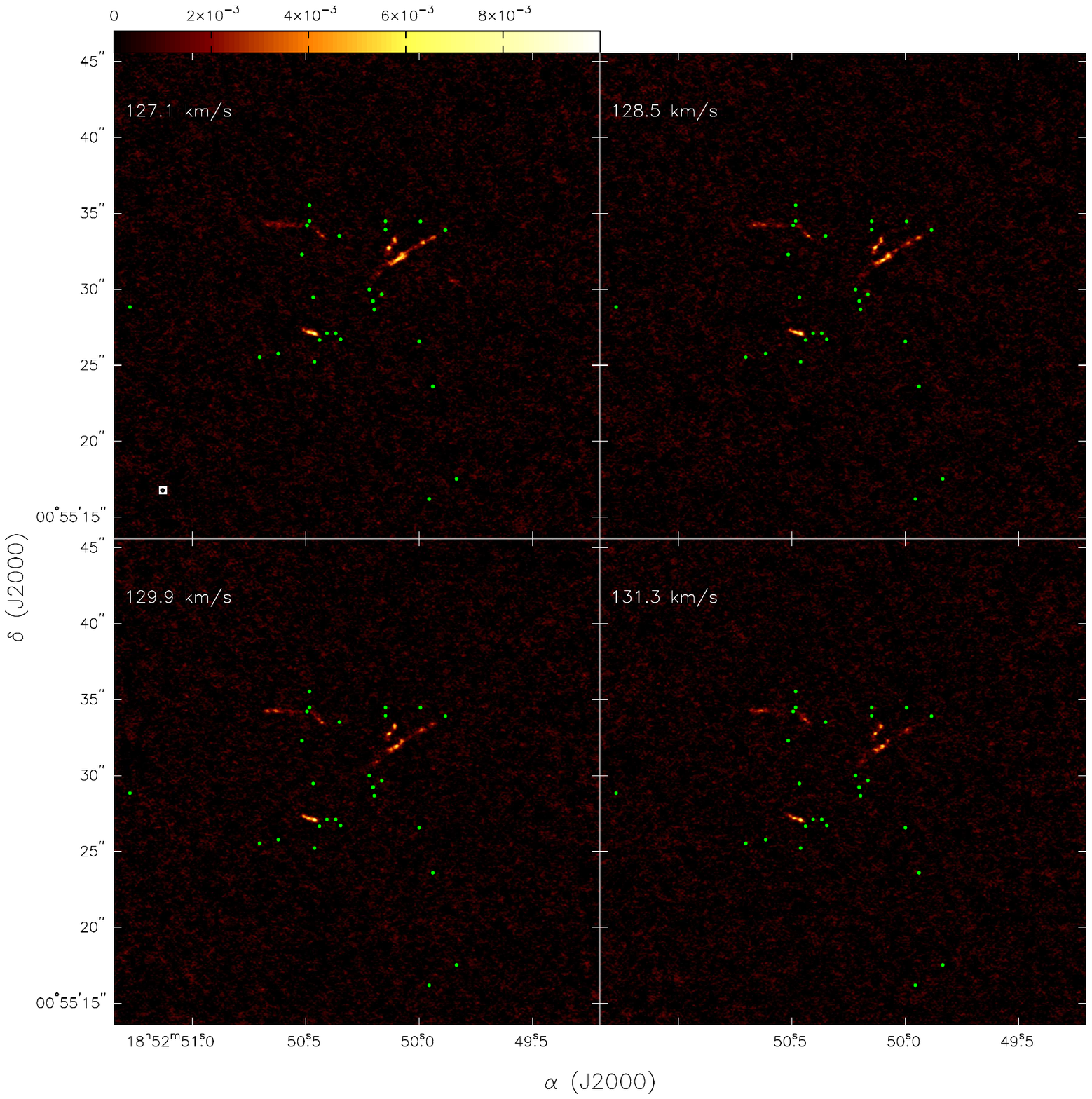}
\figsetgrpnote{Velocity channel maps of SiO 5-4.}
\figsetgrptitle{Velocity channel maps set 10.}
\figsetplot{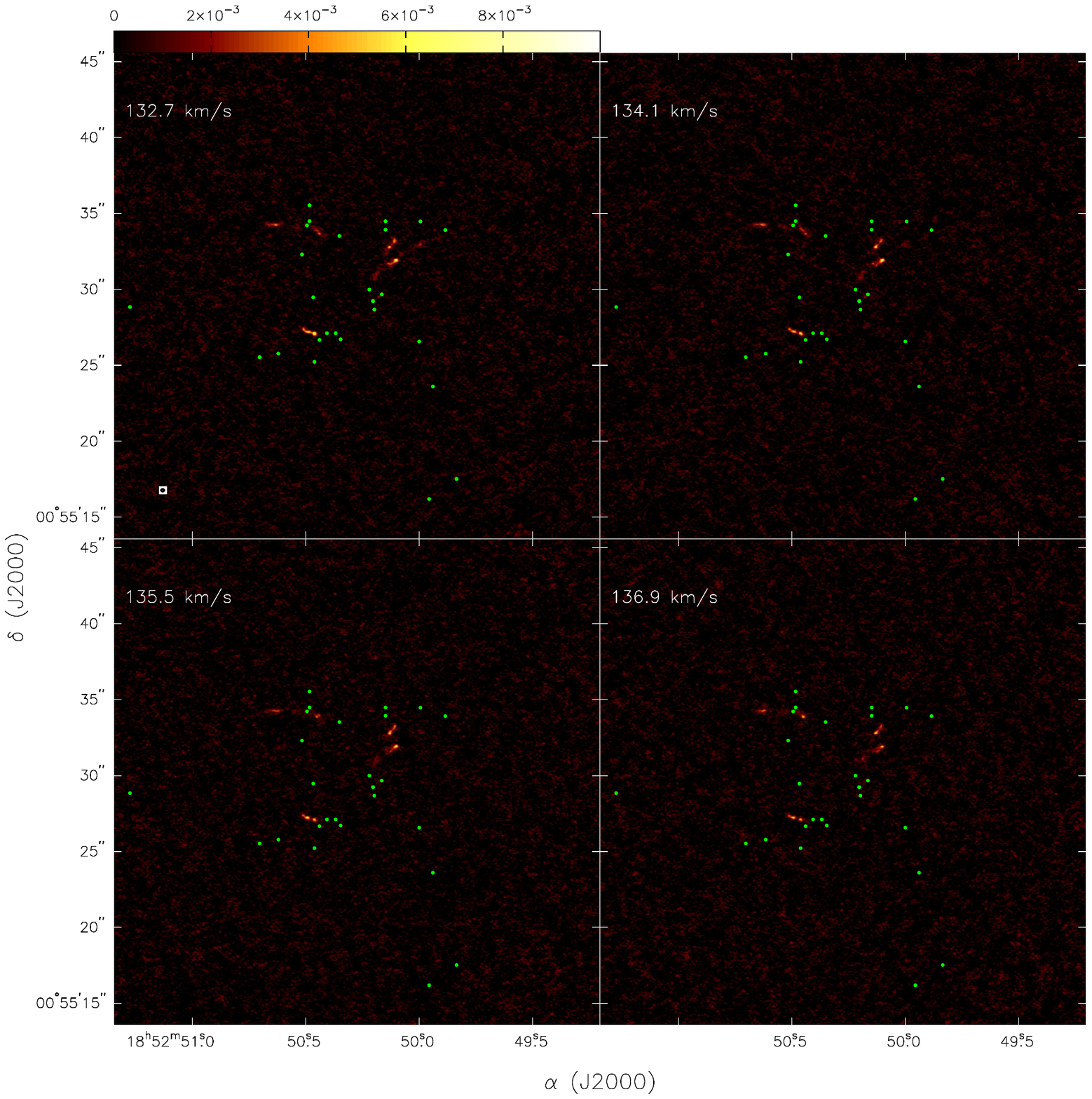}
\figsetgrpnote{Velocity channel maps of SiO 5-4.}
\figsetgrptitle{Velocity channel maps set 11.}
\figsetplot{sio_channel_g33p92_11.pdf}
\figsetgrpnote{Velocity channel maps of SiO 5-4.}
\figsetgrptitle{Velocity channel maps set 12.}
\figsetplot{sio_channel_g33p92_12.pdf}
\figsetgrpnote{Velocity channel maps of SiO 5-4.}
\figsetgrpend
\figsetend

\begin{figure}
\hspace{-1cm}
\includegraphics[trim={4cm 0.5cm 4cm 0.5cm},clip,width=23cm]{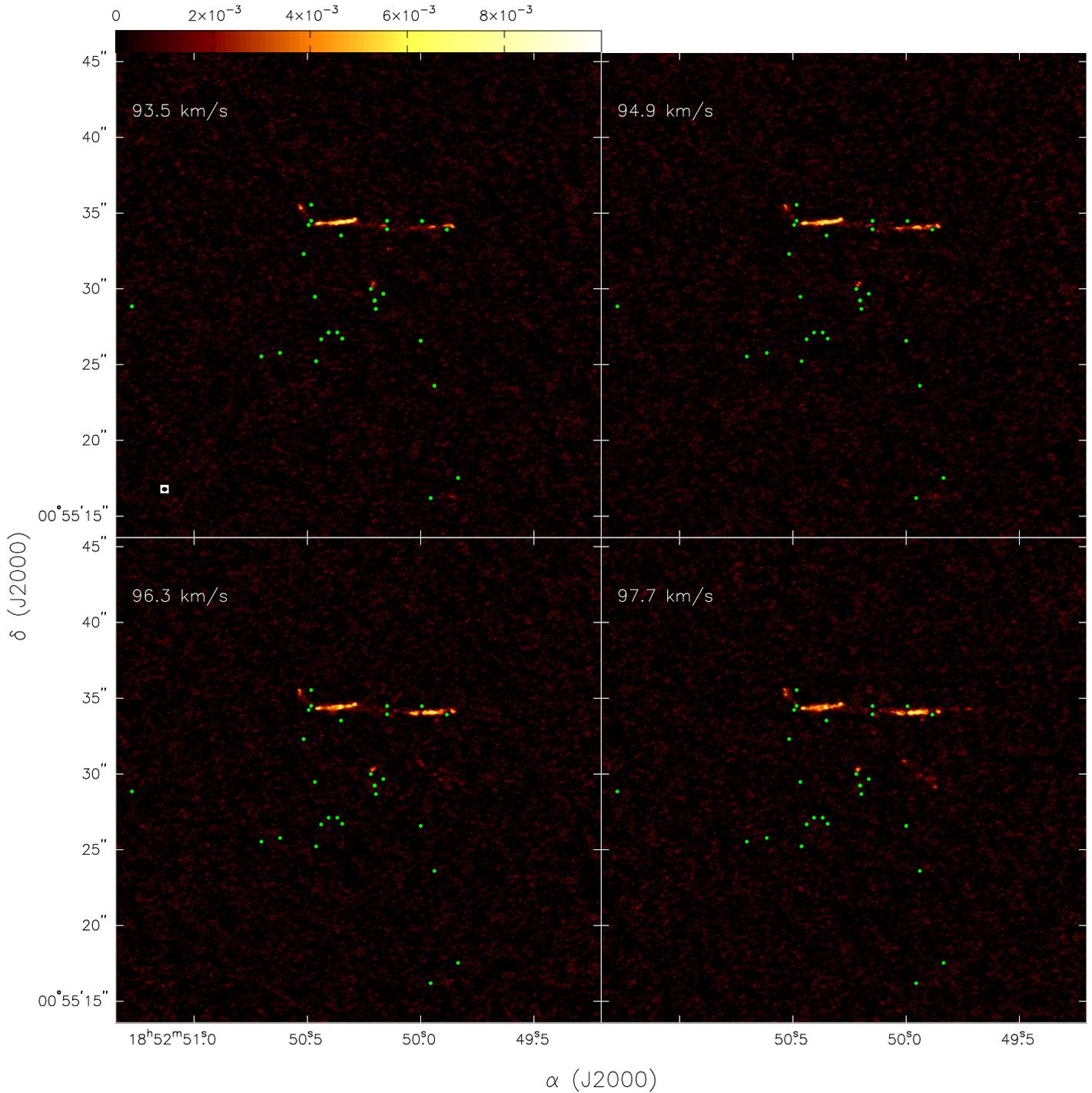}
\caption{
  The velocity channel maps of the SiO 5-4 line, in Jy\,beam$^{-1}$\,km\,s$^{-1}$ units (\beam$=$0$''$2$\times$0$''$15, P.A.=-90$^{\circ}$).
  These images are not primary beam corrected, such that it is easier to visually identify the SiO emission features close to the edge of the primary beam.
  The synthesized beam is shown in the top left panel.
  The RMS noise level of these images is 0.4 mJy\,beam$^{-1}$.
  Magenta symbols are our identified Class 0/I candidates using the {\tt dendrogram} algorithm.
  The complete figure set (10 images) is available in the online journal.
        }
   \label{fig:sio}
\end{figure}

\begin{figure}
\hspace{-1cm}
\includegraphics[trim={4cm 0.5cm 4cm 0.5cm},clip,width=23cm]{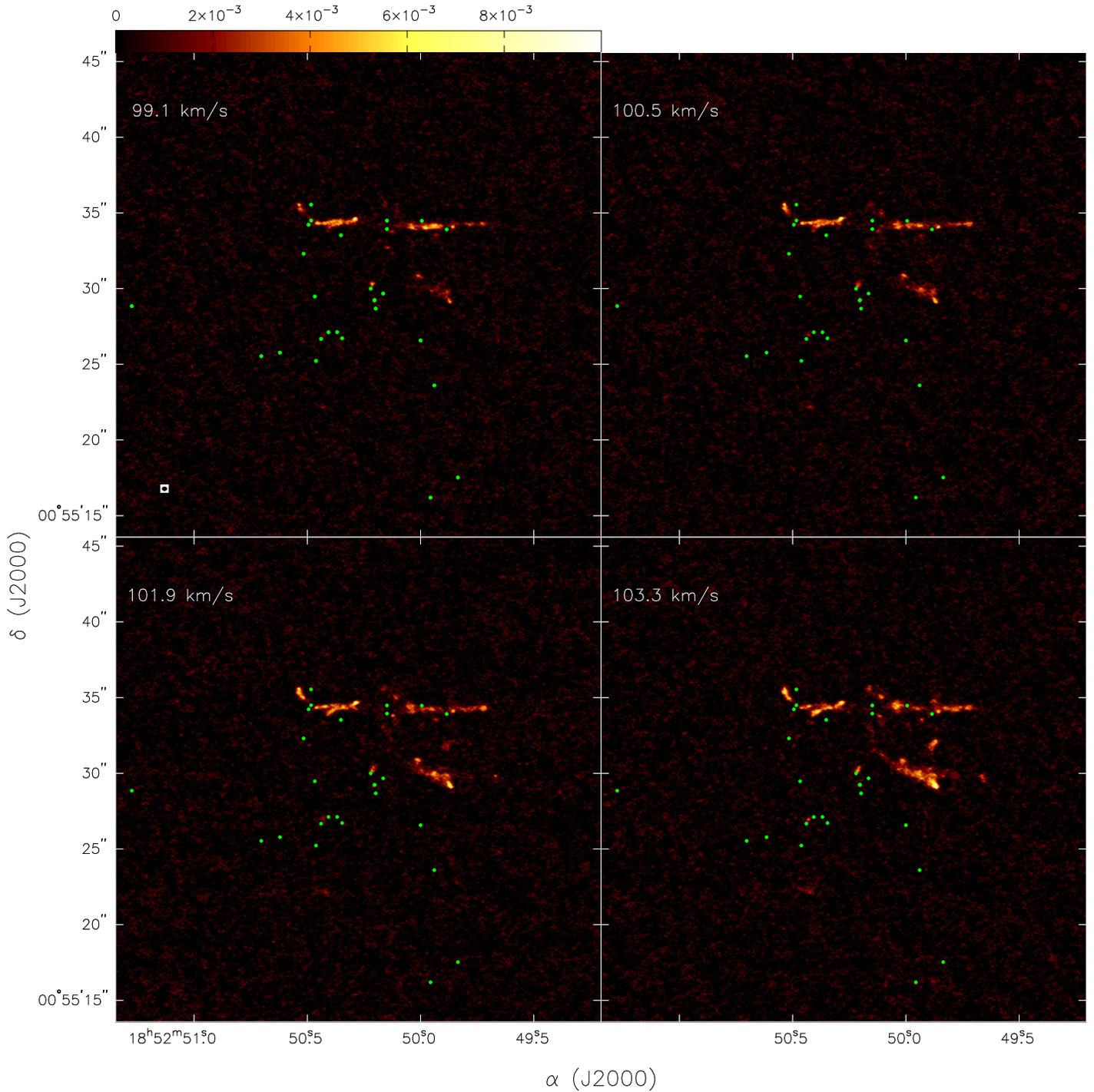}
\caption{
  The velocity channel maps of the SiO 5-4 line, in Jy\,beam$^{-1}$\,km\,s$^{-1}$ units (\beam$=$0$''$2$\times$0$''$15, P.A.=-90$^{\circ}$).
  These images are not primary beam corrected, such that it is easier to visually identify the SiO emission features close to the edge of the primary beam.
  The synthesized beam is shown in the top left panel.
  The RMS noise level of these images is 0.4 mJy\,beam$^{-1}$.
  Magenta symbols are our identified Class 0/I candidates using the {\tt dendrogram} algorithm.
  The complete figure set (10 images) is available in the online journal.
        }
   \label{fig:sio}
\end{figure}

\begin{figure}
\hspace{-1cm}
\includegraphics[trim={4cm 0.5cm 4cm 0.5cm},clip,width=23cm]{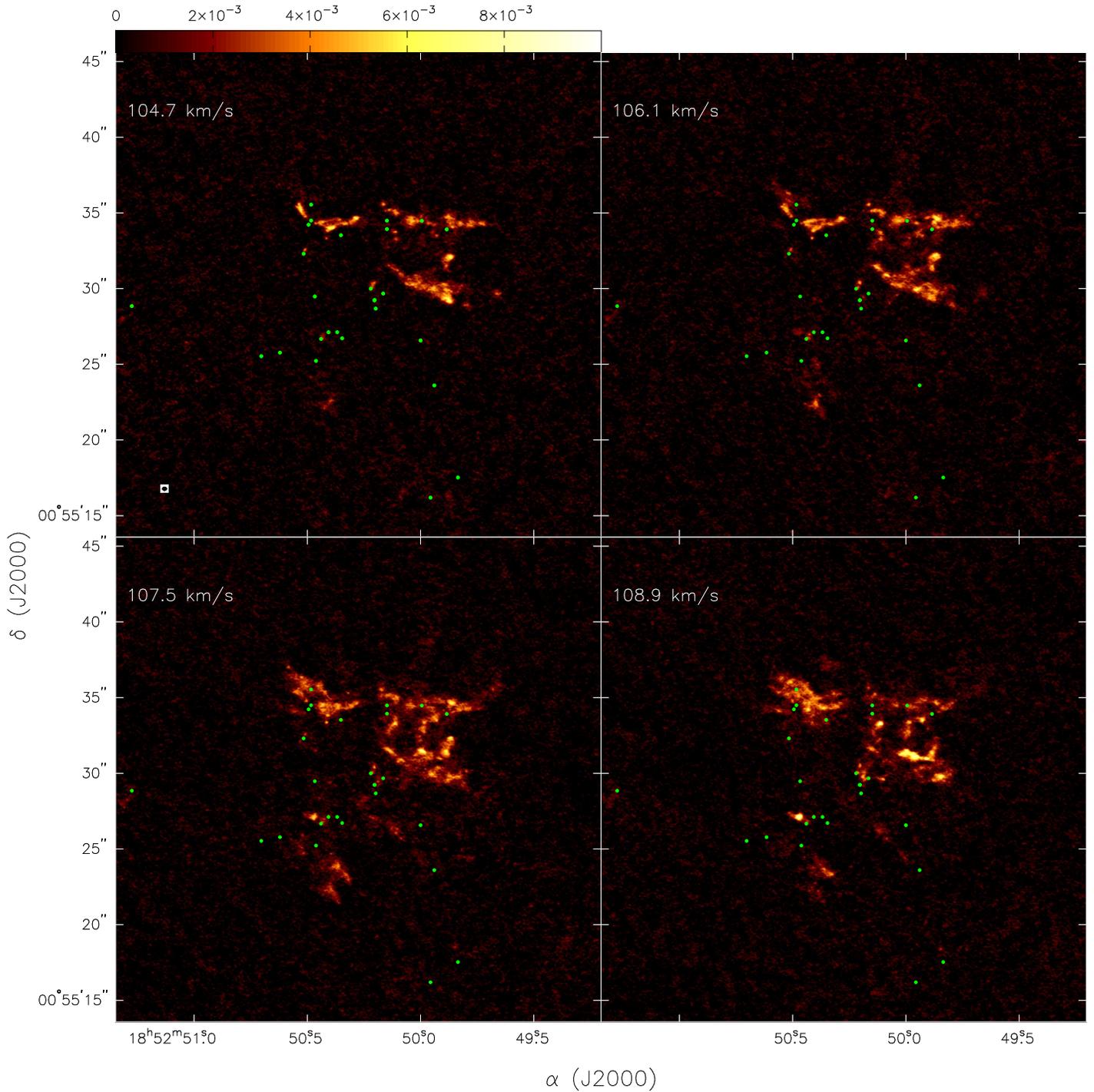}
\caption{
  The velocity channel maps of the SiO 5-4 line, in Jy\,beam$^{-1}$\,km\,s$^{-1}$ units (\beam$=$0$''$2$\times$0$''$15, P.A.=-90$^{\circ}$).
  These images are not primary beam corrected, such that it is easier to visually identify the SiO emission features close to the edge of the primary beam.
  The synthesized beam is shown in the top left panel.
  The RMS noise level of these images is 0.4 mJy\,beam$^{-1}$.
  Magenta symbols are our identified Class 0/I candidates using the {\tt dendrogram} algorithm.
  The complete figure set (10 images) is available in the online journal.
        }
   \label{fig:sio}
\end{figure}

\begin{figure}
\hspace{-1cm}
\includegraphics[trim={4cm 0.5cm 4cm 0.5cm},clip,width=23cm]{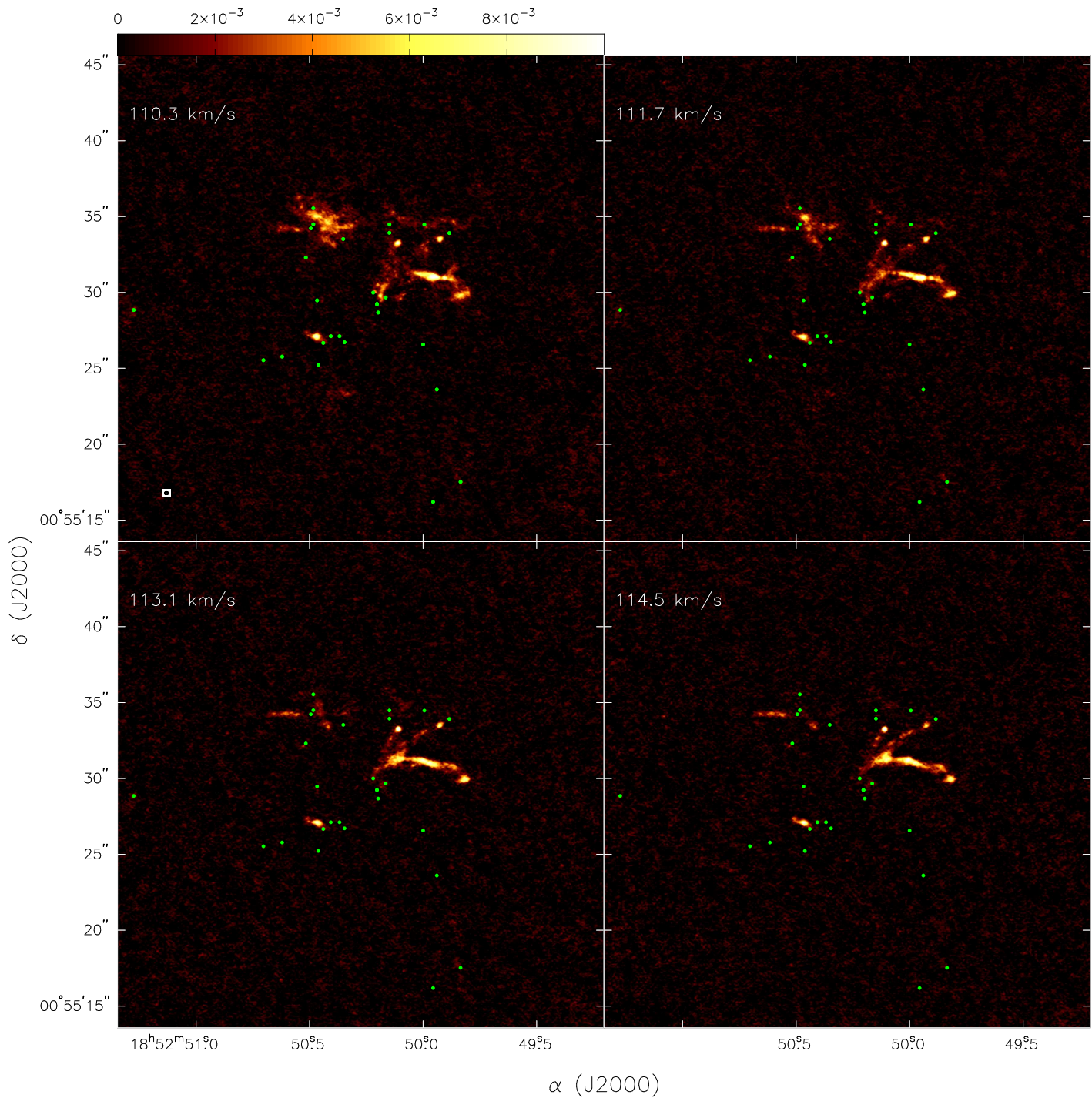}
\caption{
  The velocity channel maps of the SiO 5-4 line, in Jy\,beam$^{-1}$\,km\,s$^{-1}$ units (\beam$=$0$''$2$\times$0$''$15, P.A.=-90$^{\circ}$).
  These images are not primary beam corrected, such that it is easier to visually identify the SiO emission features close to the edge of the primary beam.
  The synthesized beam is shown in the top left panel.
  The RMS noise level of these images is 0.4 mJy\,beam$^{-1}$.
  Magenta symbols are our identified Class 0/I candidates using the {\tt dendrogram} algorithm.
  The complete figure set (10 images) is available in the online journal.
        }
   \label{fig:sio}
\end{figure}

\begin{figure}
\hspace{-1cm}
\includegraphics[trim={4cm 0.5cm 4cm 0.5cm},clip,width=23cm]{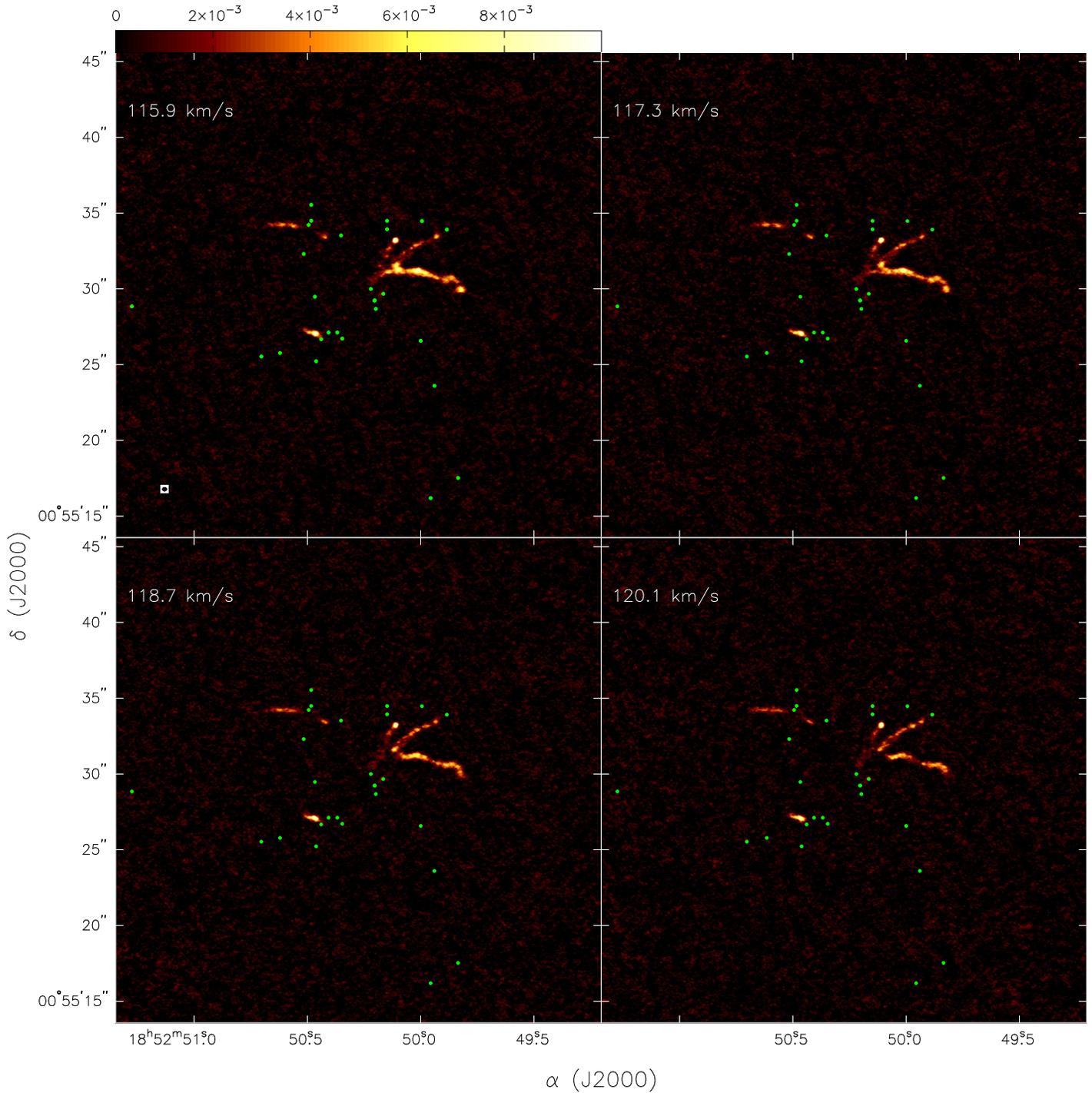}
\caption{
  The velocity channel maps of the SiO 5-4 line, in Jy\,beam$^{-1}$\,km\,s$^{-1}$ units (\beam$=$0$''$2$\times$0$''$15, P.A.=-90$^{\circ}$).
  These images are not primary beam corrected, such that it is easier to visually identify the SiO emission features close to the edge of the primary beam.
  The synthesized beam is shown in the top left panel.
  The RMS noise level of these images is 0.4 mJy\,beam$^{-1}$.
  Magenta symbols are our identified Class 0/I candidates using the {\tt dendrogram} algorithm.
  The complete figure set (10 images) is available in the online journal.
        }
   \label{fig:sio}
\end{figure}

\begin{figure}
\hspace{-1cm}
\includegraphics[trim={4cm 0.5cm 4cm 0.5cm},clip,width=23cm]{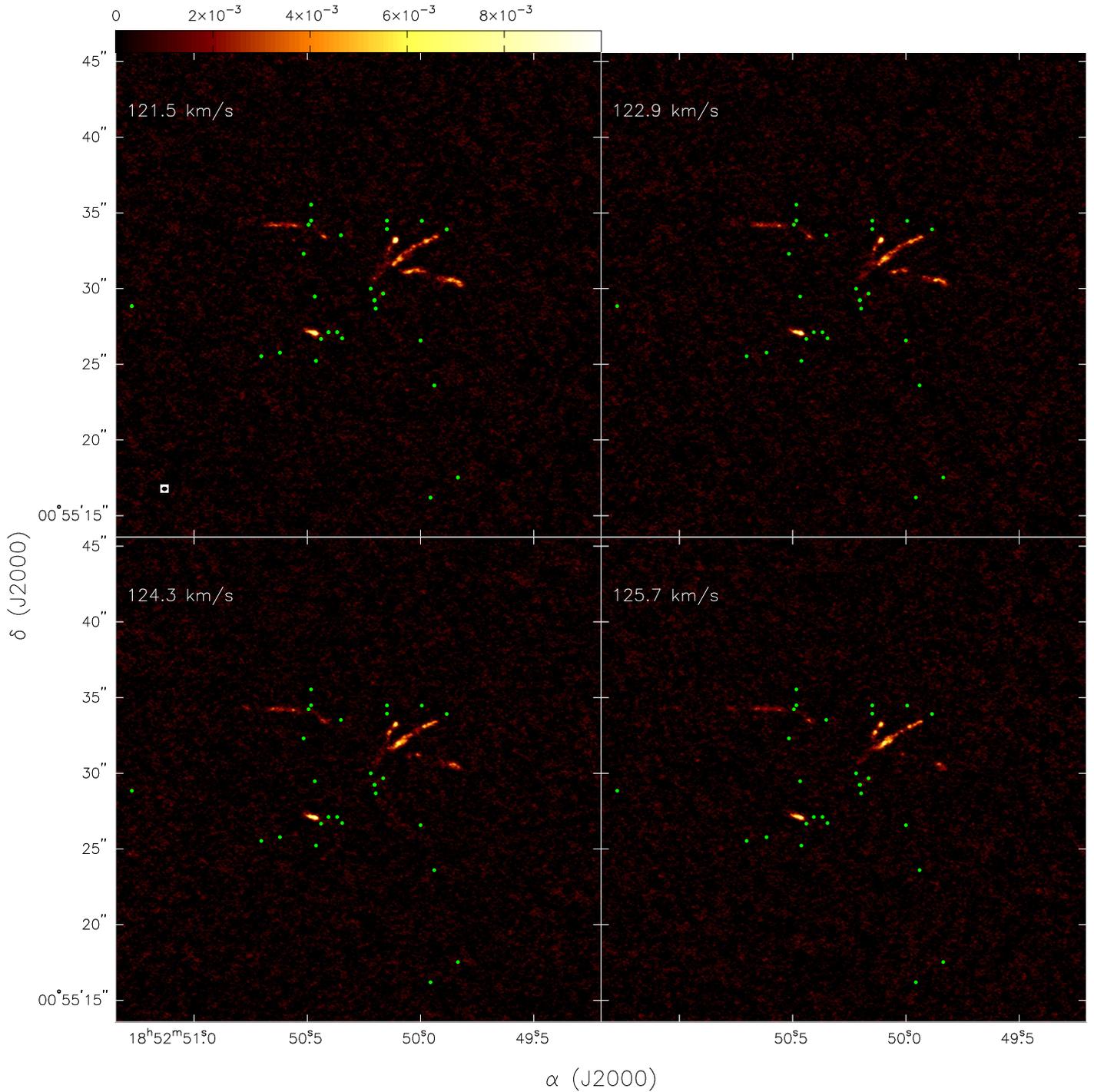}
\caption{
  The velocity channel maps of the SiO 5-4 line, in Jy\,beam$^{-1}$\,km\,s$^{-1}$ units (\beam$=$0$''$2$\times$0$''$15, P.A.=-90$^{\circ}$).
  These images are not primary beam corrected, such that it is easier to visually identify the SiO emission features close to the edge of the primary beam.
  The synthesized beam is shown in the top left panel.
  The RMS noise level of these images is 0.4 mJy\,beam$^{-1}$.
  Magenta symbols are our identified Class 0/I candidates using the {\tt dendrogram} algorithm.
  The complete figure set (10 images) is available in the online journal.
        }
   \label{fig:sio}
\end{figure}

\begin{figure}
\hspace{-1cm}
\includegraphics[trim={4cm 0.5cm 4cm 0.5cm},clip,width=23cm]{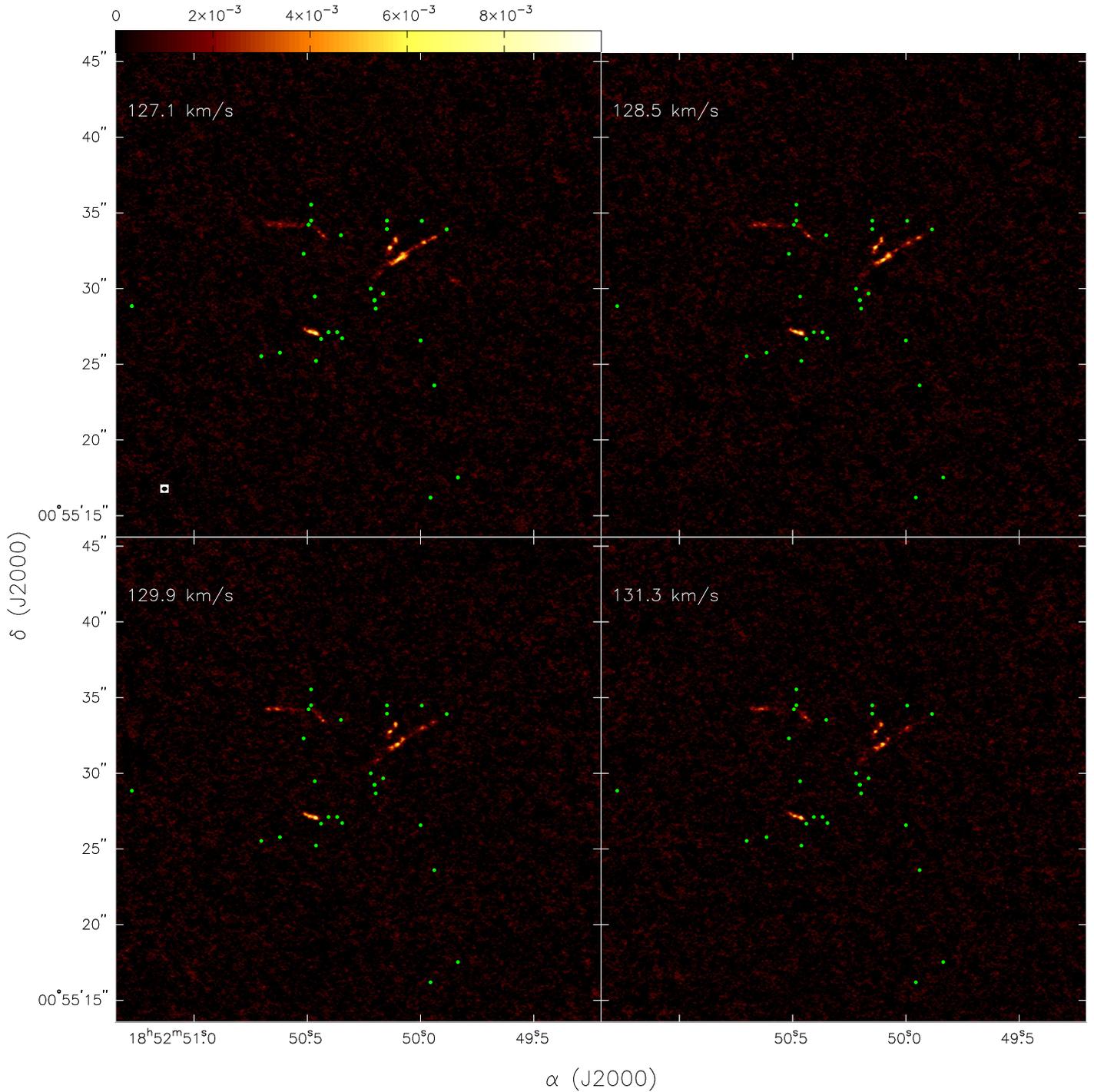}
\caption{
  The velocity channel maps of the SiO 5-4 line, in Jy\,beam$^{-1}$\,km\,s$^{-1}$ units (\beam$=$0$''$2$\times$0$''$15, P.A.=-90$^{\circ}$).
  These images are not primary beam corrected, such that it is easier to visually identify the SiO emission features close to the edge of the primary beam.
  The synthesized beam is shown in the top left panel.
  The RMS noise level of these images is 0.4 mJy\,beam$^{-1}$.
  Magenta symbols are our identified Class 0/I candidates using the {\tt dendrogram} algorithm.
  The complete figure set (10 images) is available in the online journal.
        }
   \label{fig:sio}
\end{figure}

\begin{figure}
\hspace{-1cm}
\includegraphics[trim={4cm 0.5cm 4cm 0.5cm},clip,width=23cm]{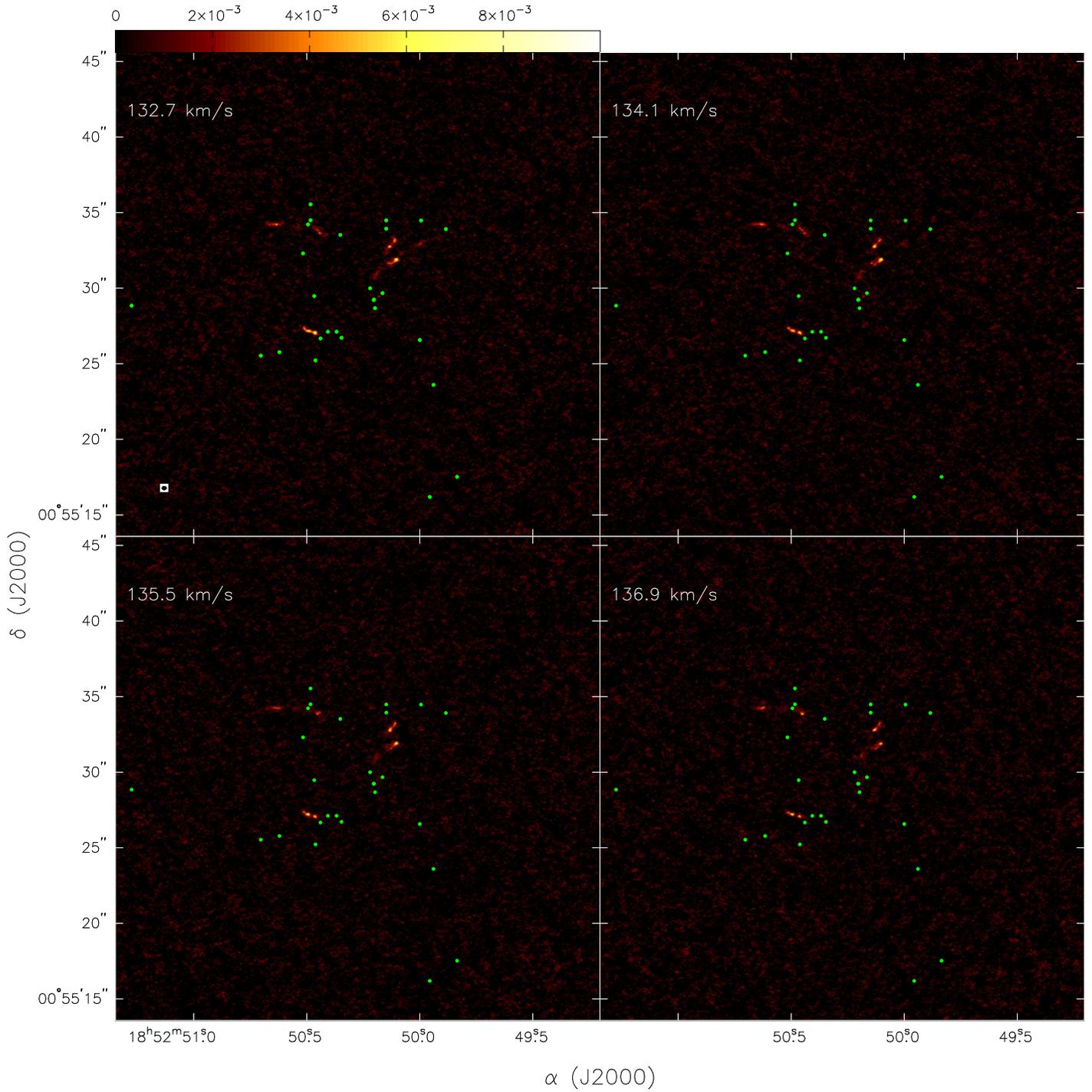}
\caption{
  The velocity channel maps of the SiO 5-4 line, in Jy\,beam$^{-1}$\,km\,s$^{-1}$ units (\beam$=$0$''$2$\times$0$''$15, P.A.=-90$^{\circ}$).
  These images are not primary beam corrected, such that it is easier to visually identify the SiO emission features close to the edge of the primary beam.
  The synthesized beam is shown in the top left panel.
  The RMS noise level of these images is 0.4 mJy\,beam$^{-1}$.
  Magenta symbols are our identified Class 0/I candidates using the {\tt dendrogram} algorithm.
  The complete figure set (10 images) is available in the online journal.
        }
   \label{fig:sio}
\end{figure}

\end{document}